\documentclass[epj]{svjour}
\usepackage{graphics}
\usepackage{amsmath}
\begin{document}
\title{Elastic theory of icosahedral
  quasicrystals - application to straight dislocations}
\author{M. Ricker\thanks{e-mail:
    mricker@itap.physik.uni-stuttgart.de}, J. Bachteler \and H.-R.
  Trebin }
\institute{Institut f\"ur Theoretische und Angewandte Physik,
  Universit\"at Stuttgart, Pfaffenwaldring 57, 70550 Stuttgart,
  Germany}
\date{Received: date / Revised version: date}
\abstract{In quasicrystals, there are not only conventional, but also phason
  displacement fields and associated Burgers vectors. We have calculated
  approximate solutions for the elastic fields induced by two-, three- and
  fivefold straight screw- and edge-dislocations in infinite icosahedral
  quasicrystals by means of a generalized perturbation method. Starting from
  the solution for elastic isotropy in phonon and phason spaces, corrections
  of higher order reflect the two-, three- and fivefold symmetry of the
  elastic fields surrounding screw dislocations. The fields of special edge 
  dislocations display characteristic symmetries also, which can be
  seen from the contributions of all orders.
  }
\maketitle
\section{Introduction}
\label{intro}
It has been demonstrated frequently (see, e.g., \cite{J1}) that in
quasicrystals, as in periodic crystals, plasticity is caused by
dislocations moving under external stresses. If the elastic fields
around dislocations are known, their influence on the lattice and the
interaction between dislocations can be calculated. In quasicrystals,
dislocations are surrounded not only by phononic, but also by
phasonic elastic fields. They are described by a generalization of the
standard elastic equations. The generalized elastic theory has
become a powerful and important tool for studying the mechanical
behaviour of quasicrystals.

The dislocation problem is solved when the medium's elastic Green's function
is available. The elastic Green's function has been established in closed form
for isotropic \cite{J2} and hexagonal \cite{J3} ordinary media, and for
pentagonal \cite{J4}, decagonal and dodecagonal \cite{J5} quasicrystals. An
approximate solution for the elastic Green's function of icosahedral
quasicrystals is given in \cite{J6}. In this paper, the results are obtained
by direct solution of the equations of balance instead by use of Green's
method.

Analytical solutions for the elastic fields around dislocations exist only for
the above mentioned cases \cite{J2,J3,J4,J5}. In this paper, we present the
solution for icosahedral quasicrystals in terms of appropriate perturbation
series.

The structure of the paper is as following. Starting from the density wave
picture, we summarize in Section \ref{sec:2} the generalized elastic theory of
icosahedral quasicrystals, including dislocation elastic theory. Section
\ref{sec:3} deals with the generalized projection method as one possibility to
solve the dislocation problem. Here we present a set of recursion formulae,
which can be used to calculate perturbation expansions of the elastic fields.
In the last section, we discuss displacement fields induced by different types
of fivefold dislocations in icosahedral quasicrystals.

Most concepts and notations in this paper are identical to those used in
\cite{J6}.

\section{Elastic theory of icosahedral quasicrystals}
\label{sec:2} 
\subsection{Some fundamentals}
\label{sec:21}

A quasicrystal is a translationally ordered structure with sharp
diffraction pattern exhibiting non-crystallographic symmetry. For this
reason, its mass density $\rho(\mathbf{x})$ can be written as a sum
over density waves:
\begin{equation}
\label{eq1}
\rho(\mathbf{x}) = \sum_{\mathbf{k} \in \mathrm{L}}
\rho_{\mathbf{k}} \mathrm{e}^{i\mathbf{k}\cdot\mathbf{x}} = 
\sum_{\mathbf{k} \in \mathrm{L}} |\rho_{\mathbf{k}}|
\mathrm{e}^{i( \mathbf{k}\cdot\mathbf{x} + \phi_{\mathbf{k}})} \, .
\end{equation}
Here, L is a module over the reciprocal quasilattice. The numbers
$\mathrm{\phi}_{\mathbf{k}}$ are the phases of the complex
coefficients $\mathrm{\rho}_{\mathbf{k}}$.

For icosahedral quasicrystals, the diffraction pattern and L display
icosahedral point symmetry. Six vectors $\mathbf{k}_{\alpha}$, $\alpha =1,
\ldots, 6,$ pointing to appropriate six of the vertices of an icosahedron can
serve as a basis of L. Phenomenological Landau theory \cite{J7,J8} shows that
there are six degrees of freedom, which can be interpreted as the phases
$\phi_{\mathbf{k}_{\alpha}}$ of the basis vectors. Thus, a frequently used
approach is the extension of the density (\ref{eq1}) to the density of a
periodic structure in six-dimensional hyperspace, which can be subjected to a
six-dimensional displacement field $\boldsymbol{\gamma}$. (\ref{eq1}) is
recovered by a cut of the hyperperiodic structure by physical space
\cite{J9}. A parametrization of the $\phi_{\mathbf{k}}$ is
\begin{equation}
\label{eq2}
\phi_{\mathbf{k}} = 
\phi_{\mathbf{k},0} 
- \boldsymbol{\kappa} \cdot 
\boldsymbol{\gamma} \, ,
\end{equation}
which involves vectors $\boldsymbol{\kappa}$ belonging to the reciprocal
hyperlattice in six dimensions. The set $\{ \boldsymbol{\kappa}_{\alpha} \}$
must be linearly independent over the real numbers and serves as basis of the
reciprocal hyperlattice, which is usually chosen to be hypercubic. Then, the
direct hyperlattice is hypercubic also, with basis vectors
$\boldsymbol{\varrho}_{\beta}$, $\beta =1, \ldots, 6,$ defined by \cite{J7}
\begin{equation}
\label{eq3}
\boldsymbol{\kappa}_{\alpha} \cdot \boldsymbol{\varrho}_{\beta} = 2
\pi \, \delta_{\alpha \beta} \, .
\end{equation}

\begin{figure}
  \vspace{0.3cm}
  \hspace{0.9cm}
\resizebox{0.35\textwidth}{!}{%
  \includegraphics {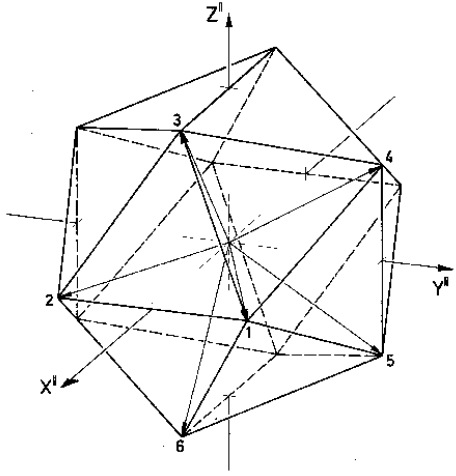} }
\caption{The projections $\mathbf{P}^{\parallel} 
\mathbf{e}_{\alpha}$ of the six 
  natural hyperspace basis vectors $\mathbf{e}_{\alpha}$ onto
  $\mathit{E}^{\parallel}$. From \cite{J9}.}
\label{fig:1}     
\end{figure}

\begin{figure}
  \vspace{0.3cm}
  \hspace{0.9cm}
\resizebox{0.35\textwidth}{!}{%
  \includegraphics {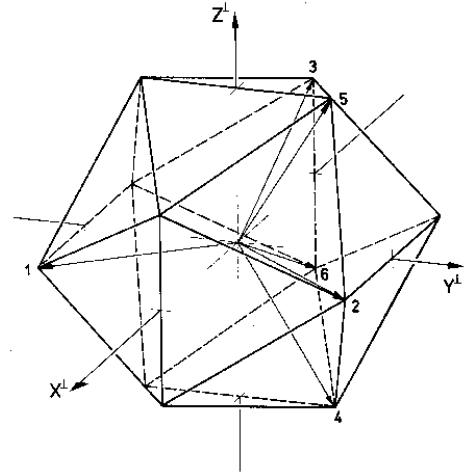} }
\caption{The projections $\mathbf{P}^{\perp} \mathbf{e}_{\alpha}$ of the six
  natural hyperspace basis vectors $\mathbf{e}_{\alpha}$ onto
  $\mathit{E}^{\perp}$. From \cite{J9}.}
\label{fig:2}    
\end{figure}

The icosahedral group Y acts on the hyperspace according to the reducible
six-dimensional representation $\Gamma^{6} = \Gamma^{3} \oplus \Gamma^{3'}$.
The hyperspace decomposes into two orthogonal, three-dimensional invariant
subspaces $\mathit{E}^{\parallel}$ and $\mathit{E}^{\perp}$, belonging to the
irreducible representations $\Gamma^{3}$ (vector representation) and
$\Gamma^{3'}$ of Y.  $\mathit{E}^{\parallel}$ is the physical or parallel
space and $\mathit{E}^{\perp}$ the perpendicular space. Two projection
operators $\mathbf{P}^{\parallel}$ and $\mathbf{P}^{\perp}$ can be applied to
hyperspace vectors to obtain their respective components in
$\mathit{E}^{\parallel}$ and $\mathit{E}^{\perp}$. This procedure is shown in
Figs. \ref{fig:1}, \ref{fig:2} for the ortho-normal natural basis $\{
\mathbf{e}_{\alpha} \}$, spanning a hypercube in six-dimensional hyperspace.

All quantities depend on the physical space coordinates
$\mathbf{x}^{\parallel} = \mathbf{x}$ only. Because of the orthogonality of
$\mathit{E}^{\parallel}$ and $\mathit{E}^{\perp}$, Eq. (\ref{eq2}) can be
written
\begin{equation}
\label{eq4}
\phi_{\mathbf{k}} = \phi_{\mathbf{k},0}
- \mathbf{k}^{\parallel} \cdot \mathbf{u} 
- \mathbf{k}^{\perp} \cdot \mathbf{w} \, . 
\end{equation}
Notations $\boldsymbol{\kappa} = \mathbf{k}^{\parallel} \oplus
\mathbf{k}^{\perp}$ and $\boldsymbol{\gamma} = \boldsymbol{\gamma}^{\parallel}
\oplus \boldsymbol{\gamma}^{\perp} = \mathbf{u} \oplus \mathbf{w}$ are used,
where $\mathbf{k}^{\parallel} = \mathbf{k} \in \mathrm{L}$.  $\mathbf{u}$ is
the ordinary phonon displacement, whose character is propagating. $\mathbf{w}$
is the phasonic displacement with diffusive character \cite{J10}. In the 
atomic picture, non-vanishing $\mathbf{w}$-fields lead to local 
rearrangements of atoms, which are called phasonic flips.


Spatially varying $\boldsymbol{\gamma}$ and $\phi_{\mathbf{k}}$, respectively,
belong to deformed states, which can be described by the elastic tensor fields
of distortion $\boldsymbol{\beta}$ and strain $\boldsymbol{\varepsilon}$
defined by
\begin{align}
\label{eq5}
\mathrm{d}\mathbf{u} & = \boldsymbol{\beta}^{u}
\mathrm{d}\mathbf{x} \, , & \mathrm{d}\mathbf{w} & =
\boldsymbol{\beta}^{w} \mathrm{d}\mathbf{x} \, , &
\boldsymbol{\beta} & = \left[
\begin{array}{c}
\boldsymbol{\beta}^{u} \\
\boldsymbol{\beta}^{w}
\end{array} 
\right], \\
\label{eq6}
\boldsymbol{\varepsilon}^{u} & = \frac{1}{2} \left(
  \boldsymbol{\beta}^{u} + \boldsymbol{\beta}^{u,t} \right)
\, , & \boldsymbol{\varepsilon}^{w} & =
\boldsymbol{\beta}^{w} \, , & \boldsymbol{\varepsilon} & =
\left[
\begin{array}{c}
\boldsymbol{\varepsilon}^{u} \\
\boldsymbol{\varepsilon}^{w}
\end{array} 
\right].
\end{align}
In case of single-valued displacement fields $\mathbf{u}$ and $\mathbf{w}$,
\begin{align}
\label{eq7}
\beta^{u}_{ij} & = \frac{\partial u_{i}}{\partial x_{j}} \, , &
\beta^{w}_{ij} & = \frac{\partial w_{i}}{\partial x_{j}} \, , \\
\label{eq8}
\varepsilon^{u}_{ij} & = \frac{1}{2} \left( \frac{\partial
    u_{i}}{\partial x_{j}} + \frac{\partial u_{j}}{\partial
    x_{i}}\right) \, , & \quad \varepsilon^{w}_{ij} & = \frac{\partial
  w_{i}}{\partial x_{j}} \, ,
\end{align}
where $i,j \in \{ 1,2,3 \}$. The six symmetric and three antisymmetric
components of $\boldsymbol{\beta}^{w}$ mix under the action of Y, and
therefore no invariants can be constructed from the symmetric or antisymmetric
components of $\boldsymbol{\beta}^{w}$ only. So due to (\ref{eq6}), the phason
strain tensor $\boldsymbol{\varepsilon}^{w}$ must be the full
$\boldsymbol{\beta}^{w}$ to contribute to the elastic energy density $F$.

The relations (\ref{eq6}) and (\ref{eq8}) hold in the linear regime
$|\frac{\partial u_{i}}{\partial x_{j}}|, |\frac{\partial w_{i}}{\partial
  x_{j}}| \ll 1$, which we are interested in. Linear elasticity is described
by an elastic energy density which is an Y-invariant quadratic form of the
components of $\boldsymbol{\varepsilon}$:
\begin{equation}
\label{eq9}
F = \frac{1}{2} \, C_{\alpha i \beta j} \, 
\varepsilon_{\alpha i} \, \varepsilon_{\beta j} = 
\frac{1}{2} \, \boldsymbol{\varepsilon} \, \mathbf{C} \,
\boldsymbol{\varepsilon} \, .
\end{equation}
Here, the Greek indices $\alpha, \beta \in \{ 1, \ldots, 6 \}$ refer to
$E^{\parallel}$ ($\alpha, \beta \in \{ 1, 2, 3 \}$) and to $E^{\perp}$
($\alpha, \beta \in \{ 4, 5, 6 \}$). This is different from the
meaning of $\alpha, \beta$ in Eq. (\ref{eq3}).

Differentiation of (\ref{eq9}) leads to the generalized Hooke's law:
\begin{align}
\label{eq10}
\sigma_{\alpha i} & = \frac{\partial F}{\partial \varepsilon_{\alpha
    i}} = C_{\alpha i \beta j} \, \varepsilon_{\beta j} \, ,
& \boldsymbol{\sigma} & = \mathbf{C} \,
\boldsymbol{\varepsilon} \, ,
\end{align}
with the generalized stress field
\begin{equation}
\label{eq11}
\boldsymbol{\sigma} = \left[
\begin{array}{c}
\boldsymbol{\sigma}^{u} \\
\boldsymbol{\sigma}^{w}
\end{array} 
\right] ,
\end{equation}
where $\boldsymbol{\sigma}^{u}=\boldsymbol{\sigma}^{u,t}$ is symmetric. 
The stress tensors $\boldsymbol{\sigma}^{u,w}$, applied to
normal vectors $\mathbf{n}$ of real or fictitious surfaces in physical
space, lead to surface forces $\mathbf{t}^{u,w} =
\boldsymbol{\sigma}^{u,w} \mathbf{n}$, which must be applied to
keep the system in balance. These forces have components in
$\mathit{E}^{\parallel}$ and $\mathit{E}^{\perp}$, respectively, and
are combined to $\mathbf{t} = \mathbf{t}^{u} \oplus \mathbf{t}^{w}$
\cite{J11}. Besides the conventional body force $\mathbf{f}^{u}$, an
analogous force $\mathbf{f}^{w}$ with direction in
$\mathit{E}^{\perp}$ has to be introduced to formulate the generalized
equations of balance consistently. We use the notation $\mathbf{f} =
\mathbf{f}^{u} \oplus \mathbf{f}^{w}$.

Hooke's law can be formulated very simply with the help of group theory. Due
to Eqs. (\ref{eq6}), $\boldsymbol{\varepsilon}^{u}$ transforms as the
representation $(\Gamma^{3} \otimes \Gamma^{3})_{sym} = \Gamma^{1} \oplus
\Gamma^{5}$ under actions of Y, whereas $\boldsymbol{\varepsilon}^{w}$
transforms as $\Gamma^{3} \otimes \Gamma^{3'} = \Gamma^{4} \oplus \Gamma^{5}$.
The transformations of the components of $\boldsymbol{\sigma}^{u,w}$ are quite
the same. With the irreducible strain components related to the coordinate
systems of Figs. \ref{fig:1} and \ref{fig:2},
\begin{equation}
\label{eq12}
\begin{aligned}
  \boldsymbol{\varepsilon}^{u}_{1} & =
  [\varepsilon^{u}_{\Gamma^{1}_{1}}]^{t} =
  [\varepsilon^{u}_{\Gamma^{1}_{1}}] \, , &
  \boldsymbol{\varepsilon}^{u}_{5} & =
  [\varepsilon^{u}_{\Gamma^{5}_{1}}, \ldots,
  \varepsilon^{u}_{\Gamma^{5}_{5}}]^{t} \, , \\
  \boldsymbol{\varepsilon}^{w}_{4} & =
  [\varepsilon^{w}_{\Gamma^{4}_{1}}, \ldots ,
  \varepsilon^{w}_{\Gamma^{4}_{4}}]^{t} \, , &
  \boldsymbol{\varepsilon}^{w}_{5} & =
  [\varepsilon^{w}_{\Gamma^{5}_{1}}, \ldots ,
  \varepsilon^{w}_{\Gamma^{5}_{5}}]^{t} \, ,
\end{aligned}
\end{equation}
(see \cite{J12} and Appendix A) and analogous vectors
containing the irreducible stresses, Hooke's law reads
\begin{equation}
\label{eq13}
\left[
\begin{array}{c}
\boldsymbol{\sigma}^{u}_{1} \\
\boldsymbol{\sigma}^{u}_{5} \\
\boldsymbol{\sigma}^{w}_{4} \\
\boldsymbol{\sigma}^{w}_{5}
\end{array} 
\right]
=
\left[
\begin{array}{cccc}
\mu_{1} & 0 & 0 & 0 \\
0 & \mu_{2} & 0 & \mu_{3} \\
0 & 0 & \mu_{4} & 0 \\
0 & \mu_{3} & 0 & \mu_{5}
\end{array} 
\right]
\left[
\begin{array}{c}
\boldsymbol{\varepsilon}^{u}_{1} \\
\boldsymbol{\varepsilon}^{u}_{5} \\
\boldsymbol{\varepsilon}^{w}_{4} \\
\boldsymbol{\varepsilon}^{w}_{5}
\end{array} 
\right] .
\end{equation}
Since only equal-indexed components of the same irreducible representation can
interact with each other, the number of independent second-order elastic
constants is restricted to five.  The elastic energy density corresponding to
(\ref{eq13}) is
\begin{align}
  F & = \frac{1}{2} \, \mu_{1} \,
  \boldsymbol{\varepsilon}^{u}_{1} \cdot
  \boldsymbol{\varepsilon}^{u}_{1} + \frac{1}{2} \, \mu_{2} \,
  \boldsymbol{\varepsilon}^{u}_{5} \cdot
  \boldsymbol{\varepsilon}^{u}_{5} + \mu_{3} \,
  \boldsymbol{\varepsilon}^{u}_{5} \cdot
  \boldsymbol{\varepsilon}^{w}_{5} \nonumber \\[1ex]
\label{eq14}
& + \frac{1}{2} \, \mu_{4} \, \boldsymbol{\varepsilon}^{w}_{4}
\cdot \boldsymbol{\varepsilon}^{w}_{4} + \frac{1}{2} \, \mu_{5}
\, \boldsymbol{\varepsilon}^{w}_{5} \cdot
\boldsymbol{\varepsilon}^{w}_{5} \, .
\end{align}
A comparison with (\ref{eq9}) yields the coefficients $C_{\alpha i \beta j}$.
In Appendix C, we discuss the conventions for independent elastic constants
used by other authors.

Mechanical stability requires $F >0$ for every $\boldsymbol{\varepsilon} \neq
0$, which is fulfilled when all eigenvalues of the elastic tensor of
(\ref{eq13}) are positive.  This leads to the following constraints on the
elastic constants: $\mu_{1} > 0$, $\mu_{2} > 0$, $\mu_{4} > 0$, $\mu_{5} > 0$,
and $\mu_{2} \mu_{5} > \mu_{3}^{2}$.

According to (\ref{eq13}) and (\ref{eq14}), the elastic constants
$\mu_{1}$ and $\mu_{2}$ describe pure phonon elasticity. Without
phason elasticity, for example on a short time scale, icosahedral
quasicrystals behave like isotropic media with the two
Lam\'{e}-constants
\begin{align}
 \label{eq15}
 \lambda & = \frac{1}{3} ( \mu_{1} - \mu_{2}) \, , & \mu & =
 \frac{1}{2} \, \mu_{2} \, .
\end{align}
They can be measured by ultrasound transmission \cite{J13}. The elastic
constants $\mu_{4}$ and $\mu_{5}$ belong to pure phason elasticity. They are
determined currently only indirectly from diffuse scattering around Bragg
peaks \cite{J14}. Values for the pho\-non-phason-coupling $\mu_{3}$ have been
calculated in computer simulations to have much smaller absolute values
than the two phononic elastic constants \cite{J15}.

Isotropic phonon elasticity in thermal equilibrium requires decoupled phonon
and phason elasticity, i.e. $\mu_{3} =0$. Isotropy in phason elasticity is
given in the spherical approximation $\mu_{4} = \mu_{5}$ discussed in
\cite{J6}, in addition to the condition $\mu_{3} =0$.

Gauss' theorem, applied to Eq. (\ref{eq10}), provides the well-known elastic
equations of balance in generalized form:
\begin{align}
\label{eq16}
\mathrm{div} \, \boldsymbol{\sigma} + \mathbf{f} & = \mathbf{0}
\, , & \frac{\partial}{\partial x_{i}} \, \sigma_{\alpha i} +
f_{\alpha} & = 0 \, ,
\end{align}
where $f_{1} = f^{u}_{x}, \ldots , f_{6} = f^{w}_{z}$. With the help
of Hooke's law and (\ref{eq8}), $\mathrm{div} \,
\boldsymbol{\sigma}$ can be written in terms of the displacement
field $\boldsymbol{\gamma}$:
\begin{align}
\label{eq17}
\mathrm{div} \, \boldsymbol{\sigma} & = \mathbf{D} (\nabla) \,
\boldsymbol{\gamma} \, , & (\mathrm{div} \,
\boldsymbol{\sigma})_{\alpha} & = D_{\alpha \beta}
(\nabla) \, \gamma_{\beta} \, ,
\end{align}
\begin{equation}
\label{eq18}
D_{\alpha \beta} (\nabla) = C_{\alpha i \beta j}
\, \frac{\partial^{2}}{\partial x_{i} \partial x_{j}} \, .
\end{equation}
Here, $\gamma_{1} = u_{x}, \ldots , \gamma_{6} = w_{z}$. According to
(\ref{eq17}), the equations of balance (\ref{eq16}) result in
\begin{equation} 
\label{eq19}
\mathbf{D} (\nabla) \, \boldsymbol{\gamma} + \mathbf{f} =
\mathbf{0} \, .
\end{equation}
Eq. (\ref{eq19}), $\mathbf{f} = \mathbf{f}^{u} \oplus \mathbf{f}^{w}$
and $\boldsymbol{\gamma} = \mathbf{u} \oplus \mathbf{w}$ imply a
decomposition of $\mathbf{D} (\nabla)$ into four $3 \times 3$ blocks:
\begin{equation}
\label{eq20}
 \mathbf{D} (\nabla) =
\left[
\begin{array}{cc}
 \mathbf{D}^{uu} (\nabla) & \mathbf{D}^{uw} (\nabla) \\ 
\mathbf{D}^{wu} (\nabla) & \mathbf{D}^{ww} (\nabla)
\end{array}
\right].
\end{equation}
Explicitly, we have\footnote{Here, some printing errors of Ref.
  \cite{J6} have been eliminated.}
\begin{equation}
\label{eq21}
\begin{aligned}
  \hspace{-1.2ex} \mathbf{D}^{uu} (\nabla) & = \mu \, \mathbf{1} \,
  \nabla^{2} + (\lambda + \mu)
  \nabla \otimes \nabla \, , \\
  \hspace{-1.2ex} \mathbf{D}^{uw} (\nabla) & = \mathbf{D}^{wu,t}
  (\nabla) \\
  \hspace{-1.2ex} & = \frac{\mu_{3}}{\sqrt{6}} \left[
\begin{array}{ccc}
 F_{1} (x,y,z) \hspace{-0.7ex} & F_{3} (x,y) 
\hspace{-0.7ex} & F_{2} (x,z) \\ F_{2} (y,x) 
\hspace{-0.7ex} & F_{1} (y,z,x) \hspace{-0.7ex} & F_{3} (y,z) \\
F_{3} (z,x) \hspace{-0.7ex} & F_{2} (z,y) \hspace{-0.7ex} & F_{1} (z,x,y)
\end{array} \right] ,
\\
\hspace{-1.2ex} \mathbf{D}^{ww} (\nabla) & = \mu_{5} \, \mathbf{1} \,
\nabla^{2} \\
& \hspace{-4.3ex} + \frac{\mu_{4} - \mu_{5}}{3} \left[
\begin{array}{ccc}
F_{4} (x,y,z) \hspace{-0.7ex} & F_{5} (x,y) \hspace{-0.7ex} & F_{5} (x,z) \\ 
F_{5} (y,x) \hspace{-0.7ex} & F_{4} (y,z,x) \hspace{-0.7ex} & F_{5} (y,z) \\
F_{5} (z,x) \hspace{-0.7ex} & F_{5} (z,y) \hspace{-0.7ex} & F_{4} (z,x,y)
\end{array} \right] .
\end{aligned}
\end{equation}
In (\ref{eq21}), we have used the abbreviations
\begin{align}
  F_{1}(a,b,c) & = - \frac{\partial^{2}}{\partial a^{2}} -
  \frac{1}{\tau} \frac{\partial^{2}}{\partial b^{2}} + \tau
  \frac{\partial^{2}}{\partial c^{2}} \, , \nonumber \\
  F_{2}(a,b) & = - \frac{2}{\tau} \frac{\partial^{2}}{\partial a
    \partial b} \, ,  \nonumber \\
  \label{eq22}
  \quad F_{3}(a,b) & = 2 \tau \,
  \frac{\partial^{2}}{\partial a \partial b} \, , \\
  F_{4}(a,b,c) & = \frac{\partial^{2}}{\partial a^{2}} + \tau^{2}
  \frac{\partial^{2}}{\partial b^{2}} + \frac{1}{\tau^{2}}
  \frac{\partial^{2}}{\partial c^{2}} \, ,   \nonumber \\
  F_{5}(a,b) & = 2 \, \frac{\partial^{2}}{\partial a \partial b} \, .
  \nonumber
\end{align}
$\tau = \frac{1}{2} (1 + \sqrt{5})$ is the Golden Mean, defined as the
positive root of the quadratic equation $x^{2} -x -1 = 0$.

\subsection{Green's function of the elastic equations of balance}
\label{sec:22}

The solution of (19) for arbitrary body forces $\mathbf{f}$ can be
obtained by calculating the integral
\begin{equation}
\label{eq23}
\boldsymbol{\gamma} (\mathbf{x})
= \int \mathbf{G} (\mathbf{x} - \mathbf{x'})
\, \mathbf{f} (\mathbf{x'}) \, \mathrm{d}^{3}x' \, ,
\end{equation}
where $\mathbf{G} (\mathbf{x} - \mathbf{x'})$ is the elastic Green's
function of icosahedral quasicrystals. The constitutive equation for
$\mathbf{G} (\mathbf{x})$ is
\begin{equation}
\label{eq24}
 \mathbf{D} (\nabla) \, \mathbf{G} (\mathbf{x})
+ \mathbf{1} \, \delta (\mathbf{x}) = \mathbf{0}\, .
\end{equation}
An approximate solution for $\mathbf{G} (\mathbf{x})$ is given in
\cite{J6}. Since (\ref{eq21}) are the components of $\mathbf{D}
(\nabla)$ for linear elasticity, the displacement field provided by
(\ref{eq23}) is the correct solution only in domains where the
linearity conditions are fulfilled.

For the purpose of this work, we only need to know the solution
$\mathbf{G}_{00} (\mathbf{x})$ for elastic isotropy in phonon and
phason spaces, defined by the conditions $\mu_{3} = 0$ and $\mu_{4} =
\mu_{5}$. With the well-known elastic Green's function for three-
dimensional, isotropic media \cite{J2} and the solution of the
fundamental Poisson's equation in three dimensions, the exact
$\mathbf{G}_{00} (\mathbf{x})$, which reflects decoupled phonon and
phason elasticity, is
\begin{equation}
\label{eq25}
\begin{aligned}
  \hspace{-0.5ex} \mathbf{G}_{00}^{uu} (\mathbf{x}) & = \frac{1}{8 \pi
    \mu (\lambda + 2 \mu)} \left[ (\lambda + 3 \mu)
    \frac{1}{|\mathbf{x}|} \, \mathbf{1} + ( \lambda + \mu )
    \frac{\mathbf{x} \otimes \mathbf{x}}{|\mathbf{x}|^{3}}
  \right] \hspace{-0.4ex} , \\
  \hspace{-0.5ex} \mathbf{G}_{00}^{ww} (\mathbf{x}) & =
  \frac{1}{4 \pi \mu_{5}|\mathbf{x}|} \, \mathbf{1} \, , \\[1.3ex]
  \mathbf{G}_{00}^{uw} (\mathbf{x}) & = \mathbf{G}_{00}^{wu}
  (\mathbf{x}) = \mathbf{0} \, .
\end{aligned}
\end{equation}

\subsection{Elastic theory of dislocations}
\label{sec:23}

A dislocation $\mathit{D}$ is characterized by its non-vanishing line integral
\begin{equation}
\label{eq26}
\oint_{\partial F_{D}} \mathrm{d} \phi_{\mathbf{k}} = 2 \pi \,
m_{\mathbf{k}} \, , \quad  m_{\mathbf{k}} = 0, \pm 1, \pm 2, \ldots \,,
\end{equation}
along any closed contour $\partial F_{D}$ surrounding the core of
$\mathit{D}$, which exists in physical space only \cite{J16,J17}.  Eq.
(\ref{eq26}) guarantees for the continuity of the mass density $\rho
(\mathbf{x})$ outside the dislocation core, as can be seen from (\ref{eq1}).
In case of periodic crystals, the lattice remains unaltered outside the
dislocation core.  Because of phasons, the same is not true in quasicrystals
\cite{J18}.

From Eq. (\ref{eq2}), condition (\ref{eq26}) can be expressed in
terms of $\boldsymbol{\kappa}$ and $\mathrm{d}
\boldsymbol{\gamma}$:
\begin{equation}
\label{eq27}
\oint_{\partial F_{D}} \boldsymbol{\kappa} \cdot  \mathrm{d} 
\boldsymbol{\gamma} = 
- 2 \pi \, m_{\mathbf{k}} \, .
\end{equation}
The six-dimensional Burgers vector is $\mathbf{b} = \mathbf{b}^{u}
\oplus \mathbf{b}^{w}$, where $\mathbf{b}^{u}$, $\mathbf{b}^{w}$ and
$\mathbf{b}$ are defined
\begin{align}
\label{eq28}
\mathbf{b}^{u} & = \oint_{\partial F_{D}} \mathrm{d} \mathbf{u} \, , &
\mathbf{b}^{w} & = \oint_{\partial F_{D}} \mathrm{d} \mathbf{w} \, , &
\mathbf{b} & = \oint_{\partial F_{D}} \mathrm{d}
\boldsymbol{\gamma} \, .
\end{align}
This allows us to rewrite (\ref{eq27}) in the form
\begin{equation}
\label{eq29}
\boldsymbol{\kappa} \cdot \mathbf{b} = \mathbf{k}^{\parallel}
\cdot \mathbf{b}^{u} + \mathbf{k}^{\perp} \cdot \mathbf{b}^{w} = - 2
\pi \, m_{\mathbf{k}} \, .
\end{equation}
As a conclusion from (\ref{eq3}) and (\ref{eq29}), $\mathbf{b}$ must be a
vector of the direct hyperlattice. The irrational orientations of
$E^{\parallel}$ and $E^{\perp}$ in hyperspace provide non-vanishing components
$\mathbf{b}^{u}$ and $\mathbf{b}^{w}$ for every dislocation.

Because of Eqs. (\ref{eq28}), the displacement fields of dislocations
are multiple-valued, and therefore not integrable. According to
potential theory, the non-integrability of Eqs.  (\ref{eq5})
corresponds to $\mathrm{curl} \, \boldsymbol{\beta}^{u,w} \neq
\mathbf{0}$. From Stokes' theorem,
\begin{align}
  \mathbf{b}^{u,w} & = \oint_{\partial F_{D}} \mathrm{d} (\mathbf{u},
  \mathbf{w}) = \oint_{\partial F_{D}} \boldsymbol{\beta}^{u,w}
  \,
  \mathrm{d} \mathbf{x} \nonumber \\
\label{eq30}
& = \int_{F_{D}} \mathrm{curl} \, \boldsymbol{\beta}^{u,w} \,
\mathrm{d} \mathbf{f} = \int_{F_{D}} \boldsymbol{\alpha}^{u,w}
\, \mathrm{d} \mathbf{f} \, ,
\end{align}
where the surface integrals extend over the surface $F_{D}$, bounded by
$\partial F_{D}$ and pierced by the dislocation $D$. Because of Eq.
(\ref{eq30}), the tensors $\boldsymbol{\alpha}^{u,w}$ are dislocation
densities in phonon and phason spaces per area of physical space. It is
\begin{align}
\label{eq31}
\boldsymbol{\alpha}^{u,w} & = \mathrm{curl} \,
\boldsymbol{\beta}^{u,w} \, , & \alpha^{u,w}_{ij} =
\epsilon_{jkl} \frac{\partial \beta_{il}^{u,w}}{\partial x_{k}} \, .
\end{align}
Eqs. (\ref{eq31}), in connection with boundary conditions to be fulfilled, are
the defining equations for $\boldsymbol{\beta}^{u,w}$. 

By means of Eqs.
(\ref{eq6}), the elastic distortion fields $\boldsymbol{\beta}^{u,w}$ lead to
the strain fields $\boldsymbol{\varepsilon}^{u,w}$.
Connected with the elastic strain fields of dislocations are stress fields
(\ref{eq10}). In the absence of body forces $\mathbf{f}$, they must
fulfill Eq. (\ref{eq16}) in the form
\begin{equation}
\label{eq32}
\mathrm{div} \, \boldsymbol{\sigma} = \mathbf{0} \, .
\end{equation}

\section{Solving the problem of a single dislocation}
\label{sec:3}

Elastic fields fulfilling Eqs. (\ref{eq10}), (\ref{eq28}) and
(\ref{eq32}) simultaneously constitute the solution of the dislocation
problem. Note that Eqs. (\ref{eq28}) for a single dislocation are a special
case of the general situation, which is characterized by an incompatibility
field \cite{J19}. The projection method \cite{J20} is a possibility to solve
the dislocation problem. A summary, containing the description of several other
methods, is given in \cite{J21}.

\subsection{The projection method}
\label{sec:31}

The idea of the projection method is the following. In a first step, an
arbitrary, multiple-valued displacement field $\hat{\boldsymbol{\gamma}} =
\hat{\mathbf{u}} \oplus \hat{\mathbf{w}}$ fulfilling (\ref{eq28}),
\begin{align}
\label{eq36}
\mathbf{b}^{u} & = \oint_{\partial F_{D}} \mathrm{d} \hat{\mathbf{u}}
\, , & \mathbf{b}^{w} & = \oint_{\partial F_{D}} \mathrm{d}
\hat{\mathbf{w}} \, , & \mathbf{b} & = \oint_{\partial F_{D}}
\mathrm{d} \hat{\boldsymbol{\gamma}} \, ,
\end{align}
has to be found. According to Eq. (\ref{eq5}), this displacement field
leads to distortions, and from (\ref{eq6}) the corresponding
eigen\-strains $\hat{\boldsymbol{\varepsilon}}^{u,w}$ can be
calculated. But the eigenstress field $\hat{\boldsymbol{\sigma}}
= \mathbf{C} \, \hat{\boldsymbol{\varepsilon}}$ will not be
compatible with the equations of balance (\ref{eq32}) in general.
There may exist a divergence $\mathrm{div} \,
\hat{\boldsymbol{\sigma}} \neq \mathbf{0}$.

In the second step, a single-valued displacement field
$\check{\boldsymbol{\gamma}} = \check{\mathbf{u}} \oplus \check{\mathbf{w}}$,
leading to strain fields $\check{\boldsymbol{\varepsilon}}^{u,w}$ and stresses
$\check{\boldsymbol{\sigma}} = \mathbf{C} \,
\check{\boldsymbol{\varepsilon}}$, due to Eqs. (\ref{eq8}) and (\ref{eq10}),
respectively, has to be found.  $\check{\boldsymbol{\gamma}}$ has to be
determined in such a way, that the divergence of the total stress field
$\boldsymbol{\sigma} = \hat{\boldsymbol{\sigma}} +
\check{\boldsymbol{\sigma}}$ vanishes:
\begin{align}
\label{eq37}
\mathrm{div} \, \boldsymbol{\sigma} = \mathrm{div} \, (
\hat{\boldsymbol{\sigma}} + \check{\boldsymbol{\sigma}}) & =
\mathbf{0} \, .
\end{align}
In this step, the stress field $\hat{\boldsymbol{\sigma}}$ is
\textit{projected} from the space of stress fields with arbitrary
divergence onto the space of stresses with vanishing divergence. The
result of this projection is the true stress field
$\boldsymbol{\sigma}$. Since the displacement field
$\check{\boldsymbol{\gamma}}$ is single-valued, we have
\begin{align}
\label{eq38}
\oint_{\partial F_{D}} \mathrm{d} \check{\mathbf{u}} & = \mathbf{0} \,
, & \oint_{\partial F_{D}} \mathrm{d} \check{\mathbf{w}} & =
\mathbf{0} \, , & \oint_{\partial F_{D}} \mathrm{d}
\check{\boldsymbol{\gamma}} & = \mathbf{0} \, .
\end{align}

Finally, the elastic fields of the dislocation are
\begin{align}
\label{eq39}
\boldsymbol{\gamma} & = \hat{\boldsymbol{\gamma}} +
\check{\boldsymbol{\gamma}} \, , & \boldsymbol{\varepsilon} &
= \hat{\boldsymbol{\varepsilon}} +
\check{\boldsymbol{\varepsilon}} \, , & \boldsymbol{\sigma}
& = \hat{\boldsymbol{\sigma}} + \check{\boldsymbol{\sigma}}
\, .
\end{align}
Eq. (\ref{eq10}) is satisfied because we have used
$(\hat{\boldsymbol{\sigma}}, \check{\boldsymbol{\sigma}}) =
\mathbf{C} \, (\hat{\boldsymbol{\varepsilon}},
\check{\boldsymbol{\varepsilon}})$. (\ref{eq28}) is fulfilled
because of Eqs. (\ref{eq36}) and (\ref{eq38}). (\ref{eq32}) is true
due to Eq. (\ref{eq37}). $\boldsymbol{\gamma}$ is
multiple-valued because $\hat{\boldsymbol{\gamma}}$ is, whereas
$\boldsymbol{\varepsilon}$ and $\boldsymbol{\sigma}$ are
single-valued.

The problem of finding the right $\check{\boldsymbol{\gamma}}$ can be solved
as described below. The equations of balance (\ref{eq19}) read for the case of
the displacement field $\check{\boldsymbol{\gamma}}$
\begin{equation}
\mathbf{D} (\nabla) \, \check{\boldsymbol{\gamma}} +
 \check{\mathbf{f}} = \mathbf{0}
\label{eq19a} \, ,
\end{equation}
where $\mathbf{D} (\nabla) \, \check{\boldsymbol{\gamma}} =
\mathrm{div} \, \check{\boldsymbol{\sigma}}$. Comparing
(\ref{eq19a}) with Eq.  (\ref{eq37}) leads to the identification
\begin{equation}
\label{eq40}
\check{\mathbf{f}} = \mathrm{div} \, \hat{\boldsymbol{\sigma}}
\end{equation}
for the right $\check{\mathbf{f}}$ to be used as fictitious body force in
(\ref{eq19a}) in order to calculate $\check{\boldsymbol{\gamma}}$. Therefore,
in the second step of the projection method, the equations of balance
\begin{equation}
\label{eq41}
\mathbf{D} (\nabla) \, \check{\boldsymbol{\gamma}} + 
\mathrm{div} \, \hat{\boldsymbol{\sigma}}
= \mathbf{0}
\end{equation}
have to be solved. We want to stress that the body force
$\check{\mathbf{f}} = \mathrm{div} \, \hat{\boldsymbol{\sigma}}$
is not a real force.

The projection method amounts to adding a singular displacement
field and a non-singular one in order to satisfy
Eq. (\ref{eq32}), which is not yet fulfilled after the first step. The
splitting up into a singular and a non-singular part and the
introduction of local axes, as done below,
has already been discussed in \cite{J16}.

\subsection{Symmetry-adapted coordinate systems}
\label{sec:32}

In this paper, we consider straight dislocations $D$ having two-,
three- and fivefold line directions. For simplicity, a
coordinate system $K^{u}_{D}$ with its $z^{u}$-axis
parallel to the respective dislocation line of $D$ should be used in
physical space. There arise certain symmetries from phasonic Burgers
vectors, which can be seen clearly when additionally choosing the
$z^{w}$-axis of the new coordinate system $K^{w}_{D}$ in phason space
parallel to that symmetry axis, which corresponds to the line
direction in physical space.

The difference between the two-, three- and fivefold dislocation lines lies in
different components of the elastic tensor $\mathbf{C}$. The components
$C_{\alpha i \beta j}$ resulting from (\ref{eq9}), (\ref{eq14}) belong to the
coordinate systems of Figs.  \ref{fig:1}, \ref{fig:2}, which are appropriate
coordinate systems for twofold dislocations. The components $C_{D,\alpha i
  \beta j}$ of the elastic tensors in symmetry-adapted coordinate systems for
the three- and fivefold dislocations are obtained by coordinate
transformations.

Consider orthogonal transformations in pho\-non and phason spaces,
\begin{align}
\label{eq42}
v^{u}_{D,i} & = \Gamma^{u}_{D,ij} \, v^{u}_{j} \, , & v^{w}_{D,i} & =
\Gamma^{w}_{D,ij} \, v^{w}_{j} \, ,
\end{align}
connecting the components of the vector $\mathbf{v}$ in the
symme\-try-adapted coordinate systems with its components belonging to
the coordinate systems of Figs.  \ref{fig:1}, \ref{fig:2}. The
transformations of the strain components are then
\begin{align}
\label{eq43}
\varepsilon^{u}_{D,ij} & = \Gamma^{u}_{D,ik} \Gamma^{u}_{D,jl}
\varepsilon^{u}_{kl} \, , & \varepsilon^{w}_{D,ij} & =
\Gamma^{w}_{D,ik} \Gamma^{u}_{D,jl} \varepsilon^{w}_{kl} \, .
\end{align}
Explicitly, we have taken
\begin{equation}
\label{eq44}
\begin{aligned}
  \boldsymbol{\mathrm{\Gamma}}^{u}_{3} & = c_{1} \left[
\begin{array}{ccc}
1 & 0 & 0 \\
0 & \tau^{2} & -1 \\
0 & 1 & \tau^{2} 
\end{array}
\right] , \quad & \boldsymbol{\mathrm{\Gamma}}^{w}_{3} & = c_{1} \left[
\begin{array}{ccc}
1 & 0 & 0 \\
0 & 1 & -\tau^{2} \\
0 & \tau^{2} & 1 
\end{array}
\right] , \\
\boldsymbol{\mathrm{\Gamma}}^{u}_{5} & = c_{2} \left[
\begin{array}{ccc}
\tau & 0 & -1 \\
0 & 1 & 0 \\
1 & 0 & \tau 
\end{array}
\right] , \quad & \boldsymbol{\mathrm{\Gamma}}^{w}_{5} & = c_{2} \left[
\begin{array}{ccc}
1 & 0 & \tau \\
0 & 1 & 0 \\
-\tau & 0 & 1 
\end{array}
\right]
\end{aligned}
\end{equation}
as orthogonal transformation matrices in (\ref{eq42}). They result
from choosing the directions $\mathbf{P}^{\parallel} (\mathbf{e}_{3} +
\mathbf{e}_{4} - \mathbf{e}_{6} )$ and $\mathbf{P}^{\parallel}
\mathbf{e}_{3}$, respectively, as dislocation lines in Fig.
\ref{fig:1}. The coefficients $c_{1}$, $c_{2}$ are
\begin{align}
\label{eq45}
c_{1} & = \frac{1}{\sqrt{1+\tau^{4}}} \, , & c_{2} & =
\frac{1}{\sqrt{1+\tau^{2}}} \, .
\end{align}
When substituting the components $\varepsilon^{u,w}_{ij}$ of Eq.  (\ref{eq9})
by components $\varepsilon^{u,w}_{D,ij}$, due to the inversion of Eqs.
(\ref{eq42}), the components $C_{D, \alpha i \beta j}$ are obtained.

The positions in phonon space are best described by cylindrical coordinates
$r$ and $\phi$, beside the cartesian coordinates $x$, $y$ in the
symmetry-adapted coordinate systems. From now on, we use the notation
$\mathbf{x} = [x,y]^{t}$ for the position perpendicular to the dislocation
lines.

\subsection{Generalized differentiation and its application to the
  projection method}
\label{sec:33}

The best $\hat{\boldsymbol{\gamma}}$ to start with in the first
step of the projection method is
\begin{equation}
\label{eq47}
\hat{\boldsymbol{\gamma}} = \frac{\mathbf{b}}{2 \pi} \, \phi \, .
\end{equation}
The resulting distortion and strain fields are
\begin{equation}
\label{eq48}
\begin{aligned}
  \hspace{-1.6ex} \hat{\boldsymbol{\varepsilon}}^{w} & =
  \hat{\boldsymbol{\beta}}^{u,w} = \frac{1}{2 \pi (x^{2}+y^{2})}
  \left[
\begin{array}{ccc}
-b^{u,w}_{1} y & \, b^{u,w}_{1} x & \, 0 \\
-b^{u,w}_{2} y & \, b^{u,w}_{2} x & \, 0 \\
-b^{u,w}_{3} y & \, b^{u,w}_{3} x & \, 0 
\end{array}
\right] , \\
\hspace{-1.6ex} \hat{\boldsymbol{\varepsilon}}^{u} & = \frac{1}{4
  \pi (x^{2}+y^{2})} \left[
\begin{array}{ccc}
-2 \, b^{u}_{1} y & \, b^{u}_{1} x -b^{u}_{2} y & \, -b^{u}_{3} y \\
b^{u}_{1} x -b^{u}_{2} y & \, 2 \, b^{u}_{2} x & \, b^{u}_{3} x\\
-b^{u}_{3} y & \, b^{u}_{3} x & \, 0 
\end{array}
\right] .
\end{aligned}
\end{equation}
The corresponding dislocation densities $\hat{\boldsymbol{\alpha}}^{u,w}$
(\ref{eq31}) are $\mathbf{0}$ for $\mathbf{x} \neq \mathbf{0}$. To consider
the behaviour at $\mathbf{x} = \mathbf{0}$, where no partial derivatives in
classical sense are defined, one has to assume all functions to be generalized
functions and calculate the generalized partial derivatives \cite{J22}. 
Note that, because of (\ref{eq47}) and (\ref{eq48}), the linear elasticity 
breaks down below a finite cut-off distance from the dislocation line. 

In case of homogeneous functions $f$ of degree $-1$ in two
variables $x$, $y$, the generalized partial derivatives are
\begin{equation}
\label{eq49}
\begin{aligned}
  \frac{\partial f}{\partial x} & = \left.  \frac{\partial f}{\partial
      x} \right|_{G} + \delta (\mathbf{x})
  \oint_{\partial G} f \, \mathrm{d}y \, , \\
  \frac{\partial f}{\partial y} & = \left.  \frac{\partial f}{\partial
      y} \right|_{G} - \delta (\mathbf{x}) \oint_{\partial G} f \,
  \mathrm{d}x \, ,
\end{aligned}
\end{equation}
where the respective first expressions 
represent certain generalized functions connected
with the classical partial derivatives. $G$ is any bounded region containing
the origin $\mathbf{x} = \mathbf{0}$.  Eqs.  (\ref{eq49}) indicate that there
may be $\delta$-functions on the dislocation lines. Their respective weight
may depend on the choice of $G$, but becomes independent of $G$ when
considering any combination of generalized partial derivatives with the total
classical derivative vanishing.

Calculating the dislocation densities (\ref{eq31}) results in
\begin{align}
  \hat{\alpha}^{u,w}_{ij} & = \delta_{j3} \, \delta (\mathbf{x})
  \oint_{\partial G} \big[ \, \hat{\beta}^{u,w}_{i1} \, \mathrm{d}x +
  \hat{\beta}^{u,w}_{i2} \,
  \mathrm{d}y \, \big] \nonumber \\
  & = \delta_{j3} \, \frac{b^{u,w}_{i}}{2 \pi} \, \delta (\mathbf{x})
  \oint_{\partial G} \big[ \, \frac{\partial \phi}{\partial x} \,
  \mathrm{d}x + \frac{\partial \phi}{\partial y} \, \mathrm{d}y \,
  \big]
  \label{eq50}  \nonumber \\
  & = \delta_{j3} \, \frac{b^{u,w}_{i}}{2 \pi} \, \delta (\mathbf{x})
  \oint_{\partial G} \mathrm{d} \phi  
   = \delta_{j3} \, b^{u,w}_{i} \, \delta (\mathbf{x}) \, .
\end{align}
Therefore, the dislocation densities are point-like,
\begin{equation}
\label{eq51}
\hat{\boldsymbol{\alpha}}^{u,w}
= \delta (\mathbf{x})
\left[
\begin{array}{ccc}
\mathbf{0}  , & \mathbf{0}  , & \mathbf{b}^{u,w} 
\end{array}
\right] .
\end{equation}
According to (\ref{eq38}),
$\check{\boldsymbol{\alpha}}^{u,w} \equiv \mathbf{0}$.

From Eq. (\ref{eq10}) and Eq. (\ref{eq48}), the stress field
$\hat{\boldsymbol{\sigma}}$ entering the projection method in the second step
is homogeneous of degree $-1$. Therefore, Eqs. (\ref{eq49}) apply when
calculating $\mathrm{div} \, \hat{\boldsymbol{\sigma}}$.

In the second step of the projection method, the solution of (\ref{eq41}) is
split into further two steps. In a first step, (\ref{eq41}) is solved for
$\mathbf{x} \neq \mathbf{0}$. This means that only the classical derivatives
$(\mathrm{div} \, \hat{\boldsymbol{\sigma}} )|_{\mathbf{x} \neq \mathbf{0}}$
enter (\ref{eq41}). Denoting the solution $\boldsymbol{\gamma}'$, Eq.
(\ref{eq41}) becomes for $\mathbf{x} \neq \mathbf{0}$
\begin{equation}
\label{eq52}
\mathbf{D} (\nabla) \, \boldsymbol{\gamma}' + 
\mathrm{div} \, \hat{\boldsymbol{\sigma}} = \mathbf{0} 
\, .
\end{equation}
The displacement field $\boldsymbol{\gamma}'$ leads to the strain
field $\boldsymbol{\varepsilon}'$ and the stress field
$\boldsymbol{\sigma}'$, due to (\ref{eq8}) and (\ref{eq10}).  The
divergence of the stress field $\hat{\boldsymbol{\sigma}} +
\boldsymbol{\sigma}'$ will vanish when considering classical
derivatives only, but may still have components proportional to
$\delta (\mathbf{x})$. In a second step, these point-like stress
sources have to be compensated. With $\mathrm{div} \,
(\hat{\boldsymbol{\sigma}} + \boldsymbol{\sigma}') = \delta
(\mathbf{x}) \, \mathbf{f}''_{L}$, Eq. (\ref{eq41}) becomes
\begin{equation}
\label{eq53} 
\mathbf{D} (\nabla) \, \boldsymbol{\gamma}'' 
+ \delta (\mathbf{x}) \, \mathbf{f}''_{L} = \mathbf{0} \, .
\end{equation}
Here, $\mathbf{f}''_{L}$ is a constant vector having the unit of a force per
length. The solution $\boldsymbol{\gamma}''$ leads to the strain field
$\boldsymbol{\varepsilon}''$ and the stress field $\boldsymbol{\sigma}''$. The
final results entering Eqs.  (\ref{eq39}) are
\begin{align}
\label{eq54}
\check{\boldsymbol{\gamma}} & = \boldsymbol{\gamma}' +
\boldsymbol{\gamma}'' \, , & \check{\boldsymbol{\varepsilon}}
& = \boldsymbol{\varepsilon}' + \boldsymbol{\varepsilon}''
\, , & \check{\boldsymbol{\sigma}} & = \boldsymbol{\sigma}'
+ \boldsymbol{\sigma}'' \, .
\end{align}

In a short excursion, we want to discuss the meaning of the divergence
$\delta (\mathbf{x}) \, \mathbf{f}''_{L}$ of the stress tensor on the
dislocation lines. According to (\ref{eq49}),
\begin{equation}
\label{eq55}
  f''_{L, \alpha} = \oint_{\partial G} \big[
  (\hat{\boldsymbol{\sigma}} +
  \boldsymbol{\sigma}')_{\alpha 1} \, \mathrm{d}y -
  (\hat{\boldsymbol{\sigma}} +
  \boldsymbol{\sigma}')_{\alpha 2} \, \mathrm{d}x \big]
  \, .
\end{equation}
(\ref{eq55}) is the $\alpha$-component of
\begin{equation}
\label{eq56}
\mathbf{t}''_{L} = \oint_{\partial G} \mathbf{t}'' \, \mathrm{d}s = 
\oint_{\partial G} (\hat{\boldsymbol{\sigma}} +
  \boldsymbol{\sigma}') \mathbf{n} \, \mathrm{d}s \, ,
\end{equation}
where $\mathbf{t}''$ is the six-dimensional surface force remaining
after the divergence of the stress tensor in terms of classical
derivatives has been brought to zero.  $\mathbf{t}''_{L}$ represents
the net force per length of the $z^{u}$-axis acting on the interior of
$G$, which must vanish \cite{J23}.

\subsection{Recursion formulae}
\label{sec:34}

Unfortunately, there is no possibility to calculate
$\check{\boldsymbol{\gamma}}$ in the second step of the projection method
exactly. We have expanded the elastic fields into perturbation series
providing an approximate solution of the dislocation problem, the zeroth order
being the solution for the case of elastic isotropy and the higher orders the
perturbation arising from the deviation from isotropy. Closely following the
projection method, the solution of the recursion formulae is the summation of
the singular displacement field of zeroth order and non-singular displacement
fields of higher order. This kind of dealing with anisotropy has already been
used in connection with Green's functions of crystals \cite{J25} and 
quasicrystals \cite{J6} and to calculate the fields of dislocations 
\cite{J25} and cracks \cite{J25a}. \cite{J23} comprises a more enclosing 
approach to the handling with elastic anisotropy.

The zeroth order is the solution for elastic isotropy ($\mu_{3} = 0$ and
$\mu_{4} = \mu_{5}$), which is known (see Appendix B). The elastic fields can
be expanded with respect to the variables $\mu_{3}$ and $\mu_{4} - \mu_{5}$.
We shall mark any physical quantity $f$ being proportional to $\mu^{m}_{3} 
(\mu_{4} - \mu_{5})^{n}$ by lower indices $mn$: $f_{mn} \sim \mu^{m}_{3}
(\mu_{4} - \mu_{5})^{n}$.

From (\ref{eq21}), it is obvious that
\begin{equation}
\label{eq57}
\mathbf{D}(\nabla) = \mathbf{D}_{00}(\nabla) + \mathbf{D}_{10}(\nabla)
+\mathbf{D}_{01}(\nabla) \, ,
\end{equation}
where according to Eqs. (\ref{eq21})
\begin{equation}
\label{eq58}
 \mathbf{D}_{00} (\nabla) =
\left[
\begin{array}{cc}
\mu \, \mathbf{1} \, \nabla^{2} +
(\lambda + \mu)
\nabla \otimes \nabla & \mathbf{0} \\ 
\mathbf{0} &  \mu_{5} \, \mathbf{1} \, \nabla^{2} 
\end{array}
\right]. 
\end{equation}
Because of (\ref{eq18}), the decomposition (\ref{eq57}) corresponds to the
following decomposition of the elastic tensor $\mathbf{C}$:
\begin{equation}
\label{eq59}
\mathbf{C} = \mathbf{C}_{00} + \mathbf{C}_{10}
+\mathbf{C}_{01} \, .
\end{equation}
For the elastic fields, the perturbation series read
\begin{align}
\label{eq60}
\boldsymbol{\gamma} & = \sum \boldsymbol{\gamma}_{mn} \, , &
\boldsymbol{\varepsilon} & = \sum
\boldsymbol{\varepsilon}_{mn} \, , & \boldsymbol{\sigma} & =
\sum \boldsymbol{\sigma}_{mn} \, .
\end{align}

The conditions, which the different orders must fulfill, are the
following. First, all displacement fields except for zeroth order must
be single-valued:
\begin{align}
\label{eq28a}
\mathbf{b} & = \oint_{\partial F_{D}} \mathrm{d}
\boldsymbol{\gamma}_{00} \, , & \mathbf{0} & = \oint_{\partial
  F_{D}} \mathrm{d} \boldsymbol{\gamma}_{mn} \quad ( mn \neq 00 )
\, .
\end{align}
This is because $\mathbf{b}$ belongs to zeroth
order. Second, Hooke's law (\ref{eq10}) must connect strains and
stresses in all orders:
\begin{equation}
\label{eq10a}
\begin{aligned}
  \boldsymbol{\sigma}_{00} & = \mathbf{C}_{00} \,
  \boldsymbol{\varepsilon}_{00} \, , \\
  \boldsymbol{\sigma}_{m0} & = \mathbf{C}_{00} \,
  \boldsymbol{\varepsilon}_{m0} + \mathbf{C}_{10} \,
  \boldsymbol{\varepsilon}_{m-1,0} \, , \\
  \boldsymbol{\sigma}_{0n} & = \mathbf{C}_{00} \,
  \boldsymbol{\varepsilon}_{0n} + \mathbf{C}_{01} \,
  \boldsymbol{\varepsilon}_{0,n-1} \, , \\
  \boldsymbol{\sigma}_{mn} & = \mathbf{C}_{00} \,
  \boldsymbol{\varepsilon}_{mn} + \mathbf{C}_{10} \,
  \boldsymbol{\varepsilon}_{m-1,n} + \mathbf{C}_{01} \,
  \boldsymbol{\varepsilon}_{m,n-1} \, .
\end{aligned}
\end{equation}
Here, the last three equations are valid for $m>0$, $n>0$ and both $m,n >
0$, respectively. Third, the divergence of the stress field of each
order must vanish:
\begin{equation}
\label{eq32a}
\mathrm{div} \, \boldsymbol{\sigma}_{mn} = \mathbf{0} \, .  
\end{equation}
The constitutive equations (\ref{eq28a}), (\ref{eq10a}) and (\ref{eq32a}) can
be solved recursively using the projection method.  At this, the zeroth order
must be treated somewhat different than higher orders.

Note that $\hat{\boldsymbol{\gamma}}$ and
$\hat{\boldsymbol{\varepsilon}}$ given by Eqs. (\ref{eq47}) and
(\ref{eq48}) belong to zeroth order, as the stress field
$\hat{\boldsymbol{\sigma}}_{00} = \mathbf{C}_{00} \,
\hat{\boldsymbol{\varepsilon}}$. $\mathrm{div} \,
\boldsymbol{\sigma}_{00} = \mathbf{0}$ must be valid, and
therefore the projection method requires the solution of the equations
of balance
\begin{equation}
\label{eq61}
\mathbf{D}_{00} (\nabla) \, 
\check{\boldsymbol{\gamma}}_{00} + 
\underbrace{\mathrm{div} \, (\mathbf{C}_{00} \, 
\hat{\boldsymbol{\varepsilon}})}_{= \check{\mathbf{f}}_{00}}
= \mathbf{0} \, .
\end{equation}
When omitting the lower indices, this is exactly the procedure of Section
\ref{sec:31}. The solution of (\ref{eq61}) leads to the strain field
$\check{\boldsymbol{\varepsilon}}_{00}$ according to Eqs.  (\ref{eq8}).
Finally, we have
\begin{equation}
\label{eq62} 
\boldsymbol{\gamma}_{00} =
\check{\boldsymbol{\gamma}}_{00} + \hat{\boldsymbol{\gamma}}
\, , \, \boldsymbol{\varepsilon}_{00} =
\check{\boldsymbol{\varepsilon}}_{00} +
\hat{\boldsymbol{\varepsilon}} \, , \,
\boldsymbol{\sigma}_{00} = \mathbf{C}_{00} (
\check{\boldsymbol{\varepsilon}}_{00} +
\hat{\boldsymbol{\varepsilon}}) \, .
\end{equation}
The full $\boldsymbol{\gamma}_{00}$ is given in Appendix B.

We demonstrate the procedure in case of higher orders for $m0 \neq 00$ only.
The approach in the other cases is completely analogous.  When calculating the
elastic fields of order $m0$, the solutions of all orders $\tilde{m}0$ with
$\tilde{m} < m$ have already been determined.  The condition for the stress
field $\boldsymbol{\sigma}_{m0}$ to be fulfilled is $\mathrm{div} \,
\boldsymbol{\sigma}_{m0} = \mathbf{0}$, where $\boldsymbol{\sigma}_{m0}$ is
determined by the second Eq.  (\ref{eq10a}). Obviously, the solution of the
equations of balance
\begin{equation}
\label{eq63}
  \mathbf{D}_{00} (\nabla) \, \boldsymbol{\gamma}_{m0} +
  \underbrace{\mathrm{div} \, (\mathbf{C}_{10} \,
  \boldsymbol{\varepsilon}_{m-1,0})}_{=\mathbf{f}_{m0}} = \mathbf{0}
\end{equation}
provides the displacement field $\boldsymbol{\gamma}_{m0}$ of
order $m0$. Eq. (\ref{eq63}) is obtained immediately by the action of
the differential operator div on both sides of the second Eq.
(\ref{eq10a}) and the fact that $\mathbf{D}_{00} (\nabla) \,
\boldsymbol{\gamma}_{m0} = \mathrm{div} \, (\mathbf{C}_{00} \,
\boldsymbol{\varepsilon}_{m0})$, with
$\boldsymbol{\gamma}_{m0}$ and
$\boldsymbol{\varepsilon}_{m0}$ connected by Eqs.  (\ref{eq8}).

Concerning the perturbation expansions of the fields
$\check{\boldsymbol{\gamma}}$, $\check{\boldsymbol{\varepsilon}}$,
$\check{\boldsymbol{\sigma}}$ and $\check{\mathbf{f}}$ introduced in Section
\ref{sec:31}, we have $\boldsymbol{\gamma}_{mn} =
\check{\boldsymbol{\gamma}}_{mn}$ and $\boldsymbol{\varepsilon}_{mn} =
\check{\boldsymbol{\varepsilon}}_{mn}$ except for zeroth order, where
(\ref{eq62}) must be applied. The relation $\boldsymbol{\sigma}_{mn} =
\check{\boldsymbol{\sigma}}_{mn}$ is true for orders $m+n>1$, whereas
$\boldsymbol{\sigma}_{00}$, $\boldsymbol{\sigma}_{10}$ and
$\boldsymbol{\sigma}_{01}$ consist of parts belonging to
$\hat{\boldsymbol{\sigma}}$ and $\check{\boldsymbol{\sigma}}$. The expansion
of $\check{\mathbf{f}}$ comprises the three terms $\check{\mathbf{f}}_{00}$,
$\check{\mathbf{f}}_{10}$ and $\check{\mathbf{f}}_{01}$ according to
(\ref{eq40}) and (\ref{eq59}), and therefore 
$\mathbf{f}_{m0}$ in (\ref{eq63})
is a completely different fictitious body force. For these reasons, the
$\check{}\,$-notation of the projection method is not appropriate in case of
Eq.  (\ref{eq63}).

As an important result of the above considerations, the perturbation
expansions of the elastic fields can be calculated by solving the equations of
balance with the differential operator $\mathbf{D}_{00} (\nabla)$ instead of
the full $\mathbf{D} (\nabla)$, a task which proves to be relatively easy. We
have performed our calculations with the help of MapleV\footnote{MapleV
  Release 4 \copyright 1981-1996 by Waterloo Maple Inc. and MapleV Release 5.1
  \copyright 1981-1998 by Waterloo Maple Inc..}, and made use of the
method described in Section \ref{sec:33}. Therefore, all displacement fields
are compounded by two parts: $\check{\boldsymbol{\gamma}}_{mn} =
\boldsymbol{\gamma}'_{mn} + \boldsymbol{\gamma}''_{mn}$ for all $mn$.

\subsection{Solving the recursion formulae}
\label{sec:35}

The calculation of the displacement fields $\boldsymbol{\gamma}'_{mn}$ can
make use of the fact that they depend on $\phi$ only:
\begin{equation}
\label{eq64}
\boldsymbol{\gamma}'_{mn}(\mathbf{x}) \equiv
\boldsymbol{\gamma}'_{mn} (\phi) \, ,
\end{equation}
with each component $\gamma_{mn,\alpha}' = c_{mn,\alpha p} \, \mathrm{e}^{i p
  \phi}$ being a real combination of complex harmonics.  Therefore, things are
much easier when expressing all derivatives (\ref{eq58}) in terms of $\partial
/ \partial r$, $\partial / \partial \phi$ and $\partial / \partial z$, with
$\partial / \partial r$ and $\partial / \partial z$ vanishing when applied to
the displacement fields (\ref{eq64}).

Transforming $\mathbf{D}_{00} (\nabla)$ yields the equations of
balance of zeroth order in the form
\begin{equation}
\label{eq69}
\begin{gathered}
  \left[
\begin{array}{lr}
-(\lambda + 2 \mu) + \mu \frac{\partial^{2}}{\partial \phi^{2}} 
\hspace{-4ex} &
- ( \lambda + 3 \mu) \frac{\partial}{\partial \phi} \\
(\lambda + 3\mu ) \frac{\partial}{\partial \phi} \hspace{-4ex} & 
- \mu + ( \lambda + 2 \mu ) \frac{\partial^{2}}{\partial \phi^{2}}
\end{array}
\right] \hspace{-1ex} 
\left[
\begin{array}{c}
u'_{r} \\[1ex] u'_{\phi}  
\end{array}
\right] +
\left[
\begin{array}{c}
\bar{f}^{u}_{r} \\[0.38ex] \bar{f}^{u}_{\phi} 
\end{array}
\right] = \mathbf{0} \, , \\
\mu \frac{\partial^{2}}{\partial \phi^{2}} u'_{z} + \bar{f}^{u}_{z} =
0 \, , \\
\mu_{5} \frac{\partial^{2}}{\partial \phi^{2}} w'_{k} +
\bar{f}^{w}_{k} = 0 \, ,
\end{gathered}
\end{equation}
where $k \in \{ x,y,z\}$. From (\ref{eq48}), all possible $(\mathrm{div} \,
\ldots)|_{\mathbf{x} \neq \mathbf{0}}$ in Eqs. (\ref{eq61}), (\ref{eq63}) and
in the analogous equations for other orders depend on $r$ like $r^{-2}$, and
in (\ref{eq69})
we use the notation $(\mathrm{div} \, \ldots)|_{\mathbf{x} \neq \mathbf{0}} =
\frac{1}{r^{2}} \bar{\mathbf{f}}$, where the length force $\bar{\mathbf{f}}$
depends on $\phi$ only. Eqs.  (\ref{eq69}) justify the ansatz (\ref{eq64}).
The order indices are omitted here.

In the second step of the calculation of $\check{\boldsymbol{\gamma}}_{mn}$,
the divergence of the stress tensor on the dislocation line must be brought to
zero. This leads to the additional displacement field
$\boldsymbol{\gamma}_{mn}''$ entering $\check{\boldsymbol{\gamma}}_{mn}$. To
perform this second step, the fields $\boldsymbol{\gamma}''_{\alpha}$
resulting from (\ref{eq53}) with the prototypical forces $\mathbf{f}''_{L
  \alpha}$ defined by $f''_{L \alpha , \beta} = \delta_{\alpha \beta}$ must be
available. We have calculated these $\boldsymbol{\gamma}''_{\alpha}$ with the
help of (\ref{eq23}) for zeroth order. They are given in Appendix B. From
there it is obvious that $\ln r$-terms and hereby dependencies on $r$ join the
displacement fields.

For the purpose of this paper, we have calculated the elastic displacement,
strain and stress fields of two- and threefold dislocations up to orders $mn$
where $m+n \le 5$, and for fivefold dislocations up to $m+n \le 15$.

For fivefold dislocations and vanishing phonon-phason-coupling, the elastic
fields are known in closed form \cite{J26}. When we neglect the
phonon-phason-coupling in our algorithm, then our analytical expressions agree
with an expansion of this closed solution into a perturbation series in
$\mu_{4}-\mu_{5}$ up to all calculated orders. But also our algorithm clearly
shows, that neither a closed solution is possible for finite
phonon-phason-coupling nor for other directions of the dislocation line.

The solution of [26] has been obtained by the more frequently used generalized
Green's method. The calculated displacements are applied to atomic models of
dislocations along a fivefold direction of i-AlPdMn to obtain
dislocation-induced atomic positions. As we shall prove in Section
\ref{sec:42}, the true fivefold symmetry of the displacement fields is
exclusively induced by the phonon-phason-coupling, but investigating the
effect of the phonon-phason-coupling on the atomic positions is beyond the
scope of this paper.

\section{Results and discussion}
\label{sec:4}

\subsection{General results}
\label{sec:41}

The strain and stress fields of all orders behave like $\frac{1}{r}$, with $r$
being the distance from the dislocation line. This is exactly compatible with
what one would expect from the exact solution, provided e.g. by the projection
method, and is also compatible with results from numerical methods.

Apart from zeroth order, the displacement fields show positive parity
perpendicular to the dislocation line:
\begin{equation}
\label{eq72}
\boldsymbol{\gamma}_{mn} (-\mathbf{x}) =
\boldsymbol{\gamma}_{mn} (\mathbf{x}) \, .
\end{equation}
In contrast to this, 
the strain and stress fields of all orders exhibit negative parity:
\begin{align}
\label{eq71}
\boldsymbol{\varepsilon}_{mn}(-\mathbf{x}) & = 
-\boldsymbol{\varepsilon}_{mn}( \mathbf{x})
\, , & \boldsymbol{\sigma}_{mn}(-\mathbf{x}) & = 
- \boldsymbol{\sigma}_{mn}(\mathbf{x}) \, .
\end{align}
These properties follow from the equations of balance (\ref{eq16}) and
(\ref{eq19}), respectively, when using the fact that all fictitious
forces occuring in the projection method have positive parity. Note that
(\ref{eq71}) guarantees for the total torque acting on the dislocation 
core to vanish.

The components $\varepsilon^{u}_{mn,33}$ and $\varepsilon^{w}_{mn,33}$ of the
strain tensors vanish. This is a consequence of the infinite geometry. Except
for the case of twofold screw dislocations, $\varepsilon^{u}_{mn,33} =
\varepsilon^{w}_{mn,33} \equiv 0$ is not compatible with $\sigma^{u}_{mn,33}
\equiv 0$ and $\sigma^{w}_{mn,33} \equiv 0$.  Therefore, stresses are present
parallel to the dislocation lines. It is not quite clear how to interpret
these statements in phason space.

The simplest dislocations are the twofold ones. Twofold edge dislocations
induce only displacements perpendicular to the dislocation line, and screw
dislocations only parallel to the dislocation line.  In case of the other
line directions, Burgers vectors and displacements perpendicular and parallel
to the dislocation lines are no longer decoupled.

For every order $mn$ where $m$ is even, the phononic displacement, strain and
stress fields depend linearly on $\mathbf{b}^{u}$ only,
whereas the
phasonic fields depend on $\mathbf{b}^{w}$ only. The contrary is
true for any order $mn$, where $m$ is odd. Additionally, we have
$\mathbf{u}_{0n} \equiv \mathbf{0}$, $\boldsymbol{\varepsilon}^{u}_{0n} \equiv
\mathbf{0}$ and $\boldsymbol{\sigma}^{u}_{0n} \equiv \mathbf{0}$ for all $n>0$.

In icosahedral quasicrystals, it is, in general, 
not possible to compute the elastic fields
belonging to a Burgers vector arising from $\mathbf{b}$ by rotation about the
dislocation line from the fields belonging to $\mathbf{b}$ by the same
rotation. This method can, for example, be applied to pentagonal
quasicrystals \cite{J4}.

\subsection{Displacement fields of fivefold dislocations}
\label{sec:42}

The following performances refer to the fivefold coordinate systems introduced
in \ref{sec:32}. The classification into phasonic edge and screw dislocations
is problematic in some respects \cite{J16}. To our opinion, this
discrimination makes sense, because only phasonic Burgers vectors which are
parallel to the $z$-axes of the coordinate systems of \ref{sec:32} do not
break the $n$-fold symmetries induced by phononic $n$-fold screw dislocations.

The higher displacement corrections shown in the next two subsections are
computed on the circle $\sqrt{x^{2} + y^{2}} = 10 \,b$ in the $xy$-plane. Here,
$b$ is the length of the different Burgers vectors we have assumed (see Eqs.
(\ref{eq73}), (\ref{eq74})). We have chosen this circle, because for our
choice of elastic constants, this is the lower limit for no strain tensor
component to exceed an absolute value of 0.1, which we regard as limit for the
validity of linear elasticity. In the diagrams with displacement components
$u_{z}$ parallel to the dislocation line, the label L stands for the arc
length of this circle.  All displacement fields are standardized to have a
vanishing net displacement on the circle, which has been achieved by adding
suitable constant displacement fields.

For the calculation of the elastic fields, the elastic con\-stants of i-AlPdMn
have been used. The phononic elastic con\-stants are $\lambda = 85
\,\mathrm{GPa}$ and $\mu \, = 65 \, \mathrm{GPa}$ \cite{J13}.  For the
phasonic elastic constants, we have taken the values $K_{1} / \mathrm{kT} =
0.1 / \mathrm{atom}$ and $K_{2} / \mathrm{kT} = -0.05 / \mathrm{atom}$ and
supposed a quenching temperature of $500^{\circ} \mathrm{C}$ \cite{J14}. This
leads to our elastic constants $\mu_{4} = 0.012 \, \mathrm{GPa}$ and $\mu_{5}
= 0.12 \, \mathrm{GPa}$. The phonon-phason-coupling has been supposed to be
$\mu_{3} = 1 \, \mathrm{GPa}$.

Note that the displacements alone are no good quantities to measure the
elastic deformation qualitatively, but the strains are. Note also, that the
phononic displacement corrections are very small compared to the phononic
Burgers vectors, and that all the corrections shown here vanish with the
phonon-phason-coupling $\mu_{3}$.

\subsubsection{Fivefold edge dislocation}
\label{sec:421}

We have chosen a six-dimensional, prototypical Burgers vector
\begin{equation}
\label{eq73}
  \mathbf{b}^{u,w} = \left[ 0, b, 0 \right]^{t} .
\end{equation}
For the length of the phononic part $\mathbf{b}^{u}$, $1 \, \mbox{\r{A}}$ is
the right order of magnitude, whereas under plastic deformation the phasonic
part $\mathbf{b}^{w}$ increases and becomes up to 100 times larger than
$\mathbf{b}^{u}$ \cite{J1}. Pure edge or screw dislocations appear very
rarely in reality. Due to the edge type of the dislocation defined by
(\ref{eq73}), the fivefold symmetry of the dislocation line is broken, as can
be seen from the displacement, strain and stress fields.

\begin{figure}
  \vspace{0.3cm}
  \hspace{0.9cm}
\resizebox{0.36\textwidth}{!}{%
  \includegraphics {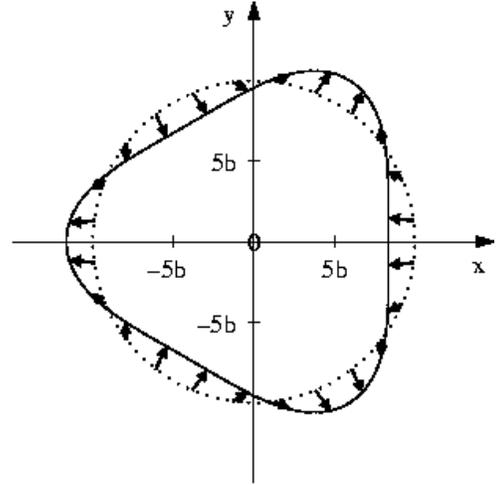} }
\caption{Phononic $xy$-displacement field of the edge dislocation
  (\ref{eq73}), order $mn = 10$. Magnification of the displacement vectors:
  $4000$.}
\label{fig:3}    
\end{figure}
\begin{figure}
  \vspace{0.3cm}
  \hspace{0.9cm}
\resizebox{0.36\textwidth}{!}{%
  \includegraphics {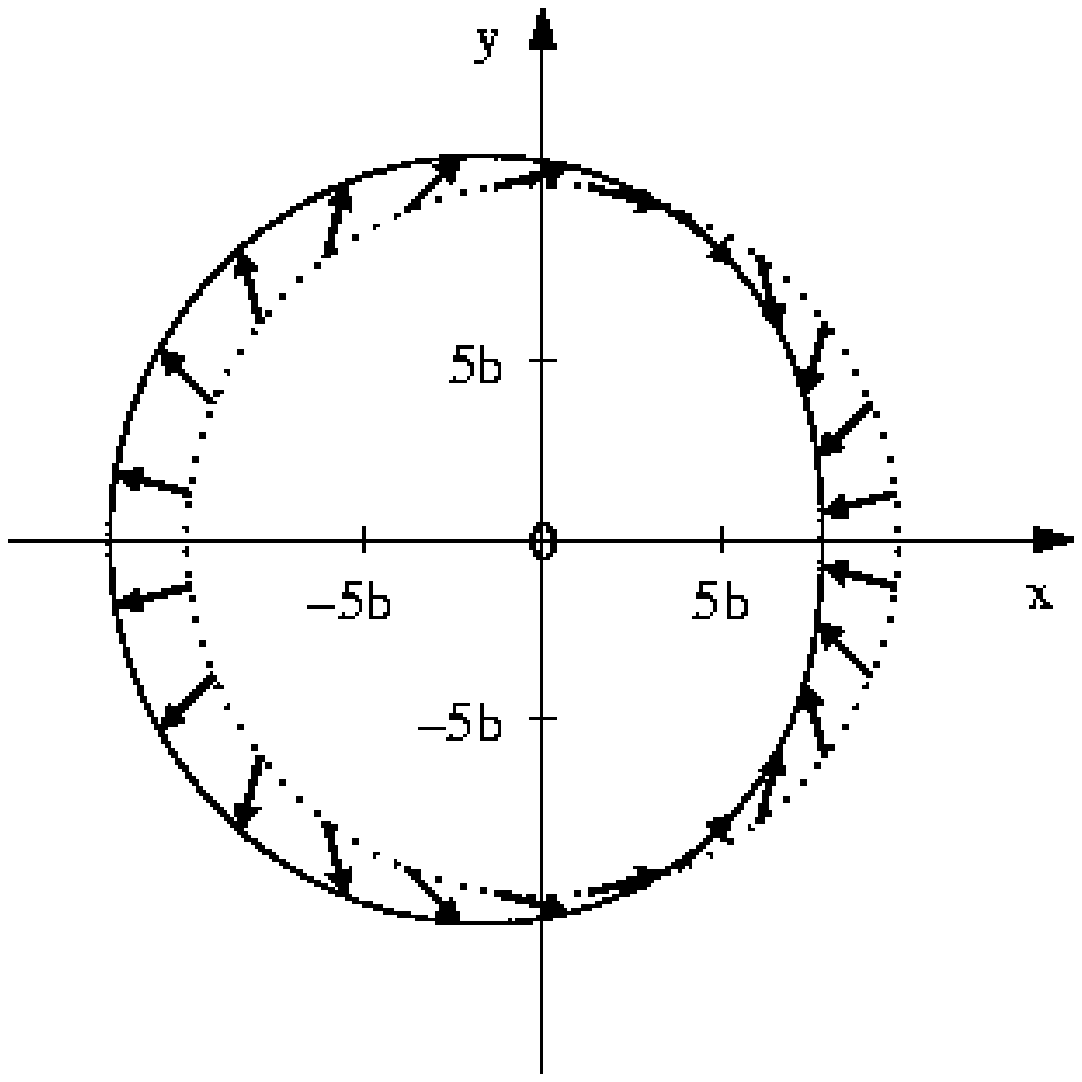} }
\caption{Phononic $xy$-displacement field of the edge dislocation
  (\ref{eq73}), order $mn = 20$. Magnification of the displacement vectors:
  $6000$.}
\label{fig:4}    
\end{figure}
\begin{figure}
  \vspace{0.3cm}
  \hspace{0.9cm}
\resizebox{0.36\textwidth}{!}{%
  \includegraphics {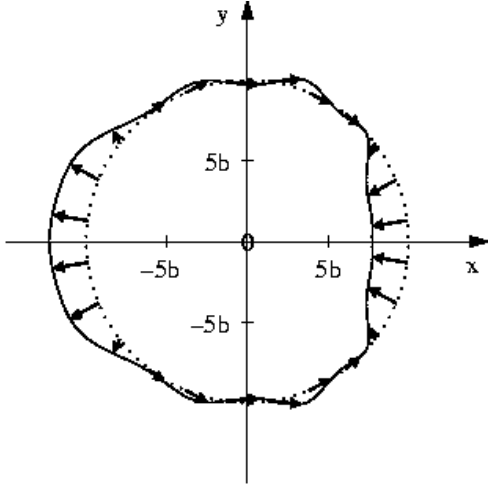} }
\caption{Phononic $xy$-displacement field of the edge dislocation
  (\ref{eq73}), all orders $mn$ with $m + n \le 15 $ except zeroth order.
  Magnification of the displacement vectors: $2500$.}
\label{fig:5}    
\end{figure}
Figs. \ref{fig:3} and \ref{fig:4} show the $xy$-components of the phononic
displacements of order $mn = 10$ and $mn = 20$. They contribute most to the
total correction of the phononic $xy$-displacement field, which is shown in
Fig. \ref{fig:5}.  Obviously, the superposition of these two low orders
already provides a good approximation for the full correction.

\begin{figure}
  \vspace{0.3cm}
  \hspace{0.9cm}
\resizebox{0.36\textwidth}{!}{%
  \includegraphics {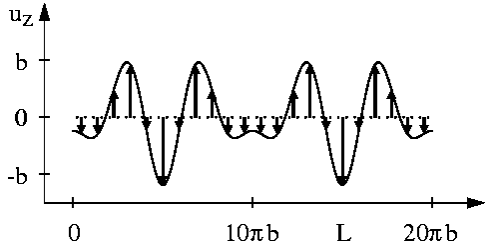} }
\caption{Phononic $z$-displacement field of the edge dislocation
  (\ref{eq73}), order $mn = 20$. Magnification of the arrows: $1200$.}
\label{fig:6}    
\end{figure}
\begin{figure}
  \vspace{0.3cm}
  \hspace{0.9cm}
\resizebox{0.36\textwidth}{!}{%
  \includegraphics {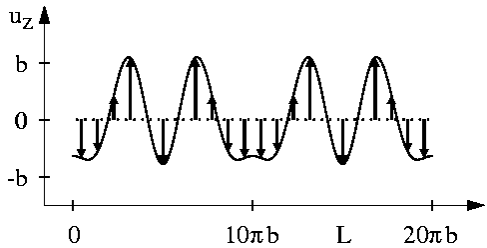} }
\caption{Phononic $z$-displacement field of the edge dislocation
  (\ref{eq73}), all orders $mn$ with $m + n \le 15 $. Magnification of the
  arrows: $2000$.}
\label{fig:7}    
\end{figure}
In Figs. \ref{fig:6} and \ref{fig:7}, the $z$-components of order $mn = 20$
and of all orders $mn$ with $m +n \le 15$ are shown. Again, the orders $mn =
10$ and $mn = 20$ give the essential contribution to the full displacement
field. In zeroth order, no $z$-component is present. The $z$-component of
order $mn = 10$ is not shown here. It proves to have a negative $\cos{2
  \phi}$-dependence, which in $2000$-fold magnification has a similar absolute
amplitude as the components of Fig. \ref{fig:6}.

The perfect twofold symmetry of the $z$-components is remarkable. This is due
to the condition (\ref{eq72}), which the displacement fields have to fulfill.
Note that the components of Figs.  \ref{fig:3}-\ref{fig:5} fulfill
(\ref{eq72}) also. In addition to this symmetry, the displacement fields show
mirror symmetry with respect to the $xz$-plane. As an evident explanation,
the Volterra cut procedure \cite{J17,J26} should be performed symmetrically
with respect to a cut in the physical $xz$-plane, so that this plane remains 
a mirror plane even after the introduction of the dislocation (see Figs. 
\ref{fig:1}, \ref{fig:2}). Our results show that the phasonic fields, which 
have to be subjected to the associated orthogonal space symmetry operation,
also have this symmetry, as it must be.

The volume contraction $\frac{ \Delta \mathrm{V}}{ \mathrm{V}} = \mathrm{tr}
\,( \boldsymbol{\varepsilon}^{u})$ and the eigenvalues of the phononic stress
tensor $\boldsymbol{\sigma}^{u}$ of zeroth order do not change substantially
from the contribution of higher orders.

When considering phasonic components, the typical displacements are more
than a hundred times larger than their phononic counterparts. The
typical phasonic strains result to be ten times larger than the phononic
strains.

We have investigated the convergence behaviour of our perturbation expansion
and also calculated the exact displacement fields by the numerical Eshelby's
method \cite{J20,J21}.  From there, the correction displacement fields of
Figs. \ref{fig:5} and \ref{fig:7} are very close to the exact ones. The same is
true for the correction displacement field for the fivefold screw dislocation,
which is discussed below.

\subsubsection{Fivefold screw dislocation}
\label{sec:422}

The Burgers vector of a screw dislocation has been chosen to be
\begin{equation}
\label{eq74}
  \mathbf{b}^{u,w} = \left[ 0, 0, b \right]^{t} \, .
\end{equation}
A screw dislocation induces no symmetry breaking parallel to the
dislocation line. Therefore, the fivefold symmetry should be
recognizable from the phononic as from the phasonic displacement
fields.

\begin{figure}
  \vspace{0.3cm}
  \hspace{0.9cm}
\resizebox{0.36\textwidth}{!}{%
  \includegraphics {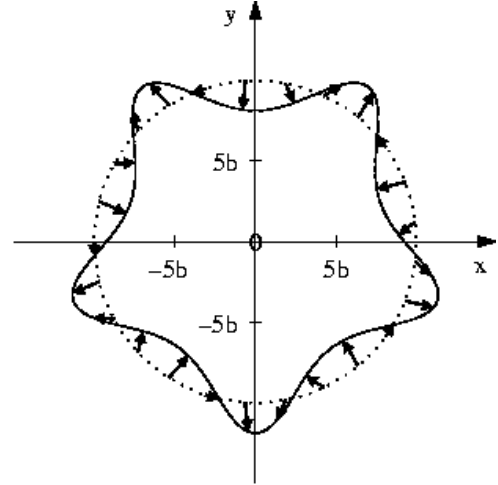} }
\caption{Phononic $xy$-displacement field of the screw dislocation
  (\ref{eq74}), order $mn = 20$. Magnification of displacement vectors:
  $2500$.}
\label{fig:8}    
\end{figure}
\begin{figure}
  \vspace{0.3cm}
  \hspace{0.9cm}
\resizebox{0.36\textwidth}{!}{%
  \includegraphics {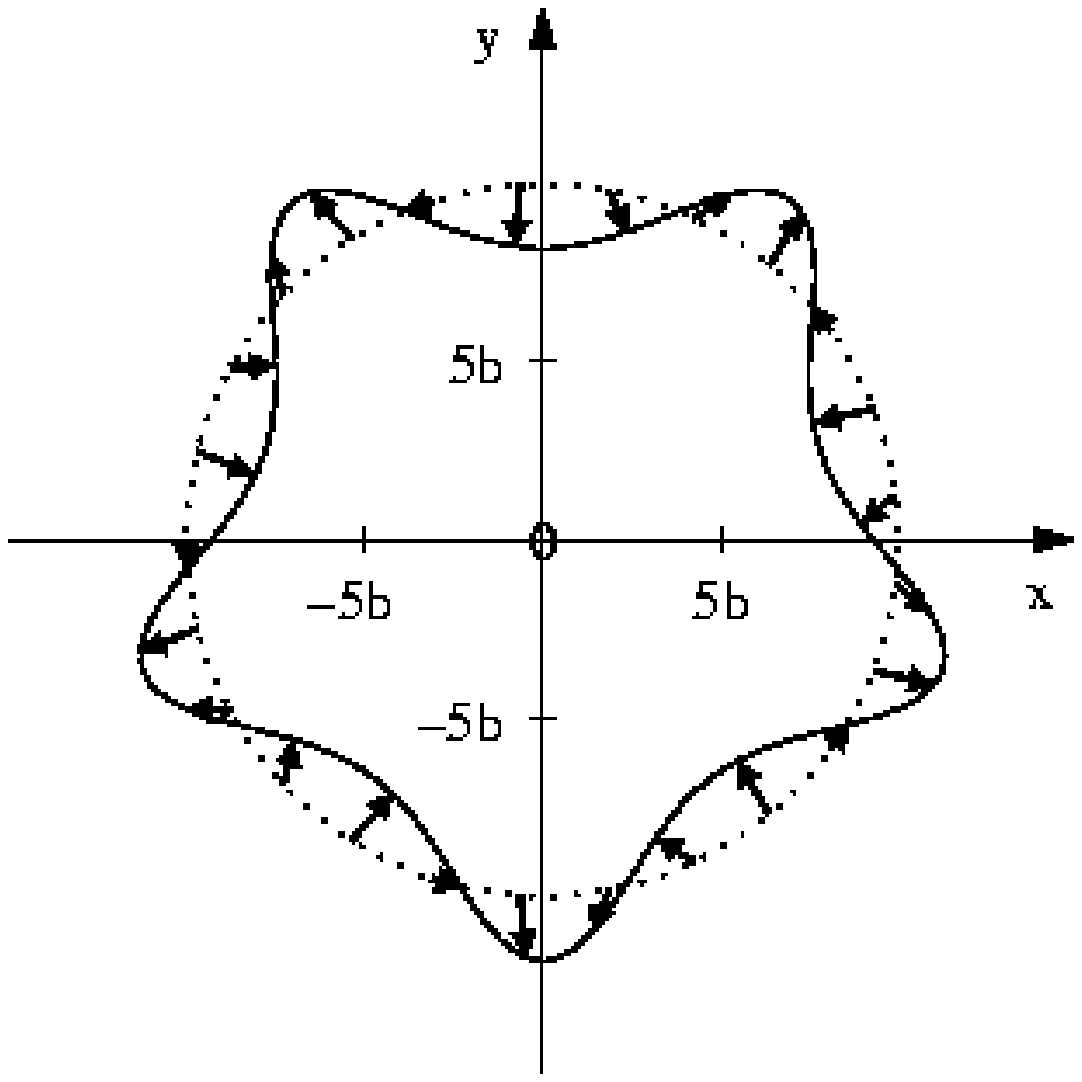} }
\caption{Phononic $xy$-displacement field of the screw dislocation
  (\ref{eq74}), all orders $mn$ with $m + n \le 15 $. Magnification of
  displacement vectors: $1500$.}
\label{fig:9}    
\end{figure}
From Fig. \ref{fig:8}, the fivefold symmetry of the phonon displacement field
is obvious. Again, $mn = 20$ is the order which provides the largest 
contribution to the complete $xy$-displace\-ment field, which is shown in Fig.
\ref{fig:9}. Although the displacement fields look qualitatively very similar,
slight differences are present in their functional dependence on $\mathbf{x}$.
No $z$-component is present in order $mn = 20$.

\begin{figure}
  \vspace{0.3cm}
  \hspace{0.9cm}
\resizebox{0.36\textwidth}{!}{%
  \includegraphics {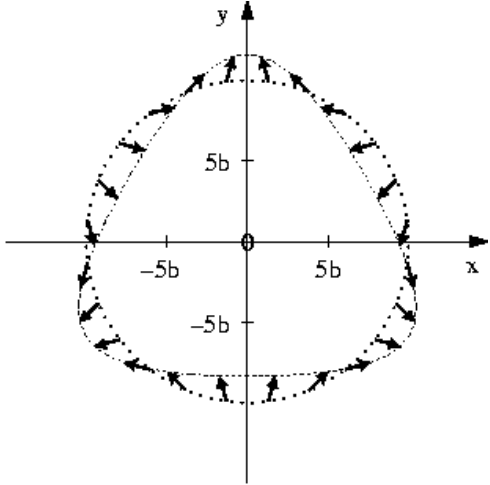} }
\caption{Phasonic $xy$-displacement field of the screw dislocation
  (\ref{eq74}), order $mn = 10$. Magnification of the displacement vectors:
  $6$.}
\label{fig:10}    
\end{figure}
In case of phason displacements, the fivefold symmetry is not easy to
recognize. Fig. \ref{fig:10} shows the phasonic displacement field of order
$mn = 10$, which provides the largest phasonic $xy$-displacements occuring in
any order. Again, no $z$-component is present. Note the small magnification,
indicating the magnitude of the phason displacements, which are appreciable
compared with the length of the phasonic Burgers vector. The position
vectors used in Fig.  \ref{fig:10} refer to the physical coordinate system of
Fig. \ref{fig:1}, but the components of the displacement vectors belong to the
phason coordinate system of Fig.  \ref{fig:2}.  Looking at the indices of the
projected hyperspace basis vectors $\mathbf{e}_{\alpha}$ in Figs. \ref{fig:1}
and \ref{fig:2}, it is clear, that a rotation about angle $\frac{2\pi}{5}$ in
physical space corresponds to a $-\frac{4\pi}{5}$-rotation in phason space.
Therefore, fivefold invariance of the displacements in Fig.  \ref{fig:10}
refers to a $-\frac{4\pi}{5}$-rotation of the vectors and a
$\frac{2\pi}{5}$-rotation of the positions.

\begin{figure}
  \vspace{0.3cm}
  \hspace{0.9cm}
\resizebox{0.36\textwidth}{!}{%
  \includegraphics {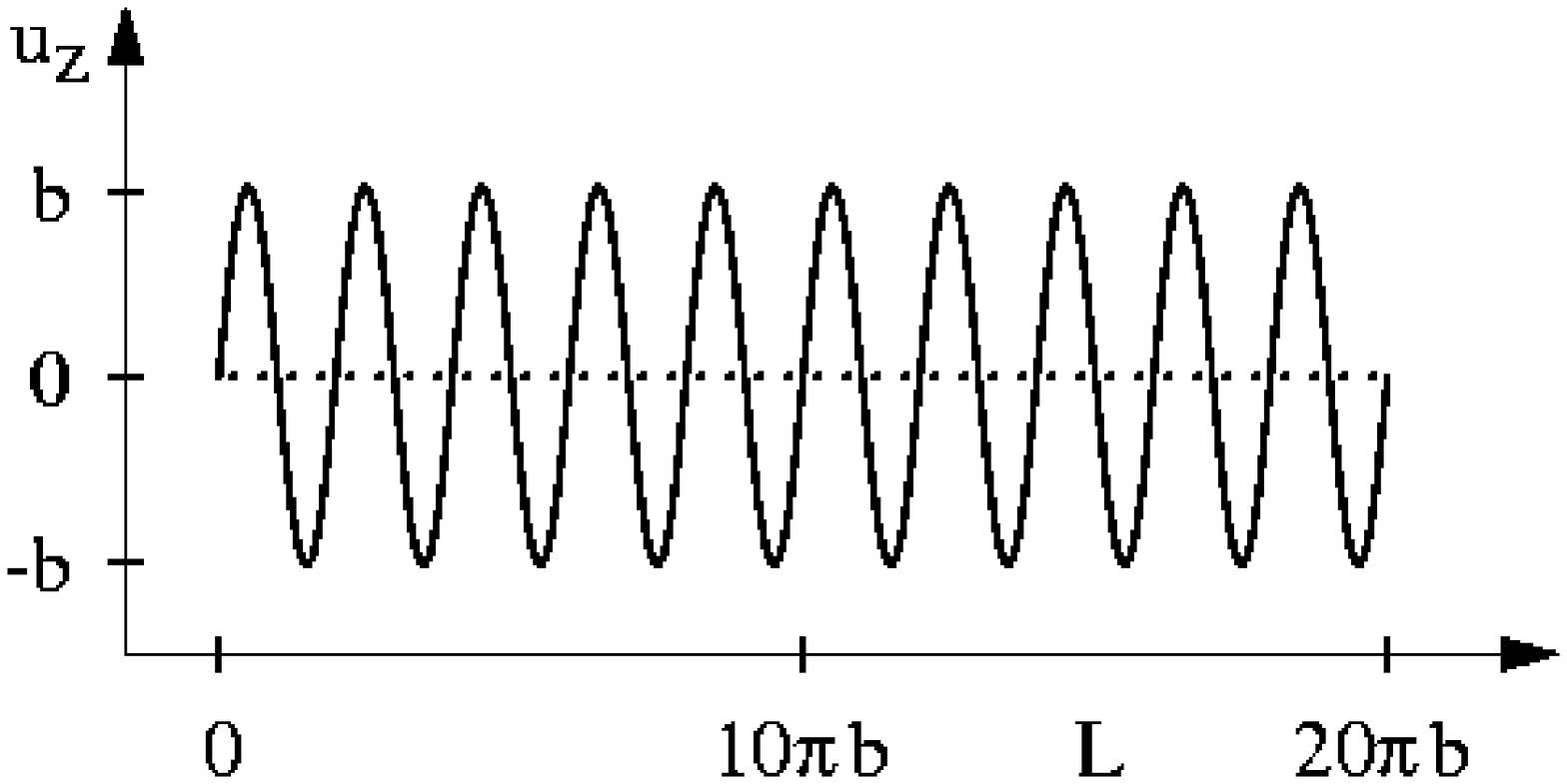} }
\caption{Phononic $z$-displacement field of the screw dislocation
  (\ref{eq74}), all orders $mn$ with $m + n \le 15 $ except zeroth order.
  Magnification of the graph: $400000$.}
\label{fig:11}    
\end{figure}
The total correction of the displacement field has a $z$-component which is
shown in Fig. \ref{fig:11}. Its tenfold symmetry is surprising, but it is the
only way for the system to obey the constraint (\ref{eq72}). The lowest orders
having such a tenfold $z$-component in phonon space are $mn = 31$ and $mn =
40$, but the $z$-component of order $mn = 40$ is about 11 times as large as in
order $mn = 31$, and provides the largest $z$-correction. The graph of Fig.
\ref{fig:11} is not exactly symmetric with respect to its extremal points.

The displacement fields exhibit no invariance under the action of an
appropriate mirror plane containing the $z$-axis, as in case of the edge
dislocation (\ref{eq73}).

The eigenvalues of the stress tensor $\boldsymbol{\sigma}^{u}$,
which are con\-stant in zeroth order, become clearly fivefold symmetric
when taking into account higher orders. The main change away from
isotropic behaviour is provided by orders $mn$ where $m + n \le 2$.
Higher orders provide also a fivefold volume contraction, whereas in
zeroth order there is none.

\section*{Acknowledgements} 

We thank \textsc{M. Boudard} and \textsc{A. L\'{e}toublon} for helpful
discussions on phasonic elastic constants. \textsc{M. R.} wants to
thank \textsc{J. Roth} and \textsc{H. Stark} for reading proofs of the first
manuscripts of this paper.

\section*{Appendix A: Icosahedral irreducible strains}

The icosahedral irreducible strains given here are taken from
\cite{J12}.
\begin{equation}
\begin{aligned}
  \boldsymbol{\varepsilon}^{u}_{1} & = \frac{1}{\sqrt{3}} \,
  (\varepsilon^{u}_{11} + \varepsilon^{u}_{22} +
  \varepsilon^{u}_{33}) \, , \\
  \boldsymbol{\varepsilon}^{u}_{5} & = \left[
  \begin{array}{c}
  \frac{1}{2\sqrt{3}} \, (- \tau^{2} \varepsilon^{u}_{11} + 
  \frac{1}{\tau^{2}} \varepsilon^{u}_{22} +
  (\tau+\frac{1}{\tau}) \varepsilon^{u}_{33}) \\
  \frac{1}{2} ( \frac{1}{\tau} \varepsilon^{u}_{11} - 
  \tau \varepsilon^{u}_{22} +
  \varepsilon^{u}_{33}) \\
  \sqrt{2} \, \varepsilon^{u}_{12} \\
  \sqrt{2} \, \varepsilon^{u}_{23} \\
  \sqrt{2} \, \varepsilon^{u}_{31}
  \end{array}
\right] ,\\
\boldsymbol{\varepsilon}^{w}_{4} & = \frac{1}{\sqrt{3}} \left[
  \begin{array}{c}
  \varepsilon^{w}_{11} + 
  \varepsilon^{w}_{22} + \varepsilon^{w}_{33} \\
  \frac{1}{\tau} \varepsilon^{w}_{21} + \tau \varepsilon^{w}_{12} \\
  \frac{1}{\tau} \varepsilon^{w}_{32} + \tau \varepsilon^{w}_{23} \\
  \frac{1}{\tau} \varepsilon^{w}_{13} + \tau \varepsilon^{w}_{31} 
  \end{array}
\right] , \\
\boldsymbol{\varepsilon}^{w}_{5} & = \frac{1}{\sqrt{6}} \left[
  \begin{array}{c}
  \sqrt{3} \, ( \varepsilon^{w}_{11} - 
  \varepsilon^{w}_{22} ) \\
  \varepsilon^{w}_{11} + 
  \varepsilon^{w}_{22} - 2 \, \varepsilon^{w}_{33} \\
  \sqrt{2} \, ( \tau \varepsilon^{w}_{21} - 
  \frac{1}{\tau} \varepsilon^{w}_{12}) \\
  \sqrt{2} \, ( \tau \varepsilon^{w}_{32} - 
  \frac{1}{\tau} \varepsilon^{w}_{23} ) \\
  \sqrt{2} \, ( \tau \varepsilon^{w}_{13} - 
  \frac{1}{\tau} \varepsilon^{w}_{31}) 
  \end{array}
\right] . 
\end{aligned}
\end{equation}

\section*{Appendix B: Elastic displacement fields of zeroth order}

From (\ref{eq47}), (\ref{eq61}) and (\ref{eq62}),
the displacement field of zeroth order is
\begin{align}
  \boldsymbol{\gamma}_{00} = \frac{1}{2 \pi} \bigg[ \, & b^{u}_{x} \, \Big\{
  \phi + \frac{ ( \lambda +\mu ) \, xy}{ ( \lambda + 2 \mu )(x^{2} + y^{2}) }
  \Big\}
  \nonumber \\
  + \, & b^{u}_{y} \, \Big\{ \frac{ \mu }{\lambda + 2 \mu } \ln
  \frac{r}{r_{0}} + \frac{ ( \lambda +\mu ) \, y^{2} }{( \lambda + 2 \mu
    )(x^{2} + y^{2}) } \Big\} \, ,
  \nonumber \\
  & b^{u}_{y} \, \Big\{ \phi - \frac{ ( \lambda + \mu ) \, xy}{( \lambda + 2
    \mu )(x^{2} + y^{2}) } \Big\}
  \nonumber \\
  - \, & b^{u}_{x} \, \Big\{ \frac{ \mu }{ \lambda + 2 \mu } \ln
  \frac{r}{r_{0}} + \frac{ ( \lambda +\mu ) \, x^{2}}{( \lambda + 2 \mu
    )(x^{2} + y^{2}) } \Big\} \, ,
  \nonumber \\
   \label{eqA11}
  & b^{u}_{z} \, \phi \, , \, b^{w}_{x} \, \phi \, , \, b^{w}_{y} \,
  \phi \, , \, b^{w}_{z} \, \phi \, \bigg]^{t} \, .
\end{align}
Here, $\gamma_{00,1} = u_{00,x}, \ldots , \gamma_{00,6} = w_{00,z}$.  This
displacement field is identical to the solution presented in \cite{J26} when
$K_{2} = 0$. $K_{2} = 0$ corresponds to identical elastic constants $\mu_{4} =
\mu_{5}$, what is obvious from (\ref{eqA21}). The factor $r_{0}$ is introduced
for dimensional reasons, and sometimes it is interpreted as inner cut-off for
the linear elasticity \cite{J4}.

The application of the projection method requires the knowledge of the
solutions of the elastic equations of balance
\begin{equation}
\label{eq53a} 
\mathbf{D}_{00} (\nabla) \, \boldsymbol{\gamma}''_{\alpha} 
+ \delta (\mathbf{x}) \, \mathbf{f}''_{L \alpha} = \mathbf{0} \, ,
\end{equation}
where the forces $\delta (\mathbf{x}) \, \mathbf{f}''_{L \alpha}$, $f''_{L
  \alpha, \beta} = \delta_{\alpha \beta} $, $\alpha, \beta = 1, \ldots , 6,$
represent point-like stress sources on the dislocation line (these
prototypical forces are not body forces!). We have obtained the resulting
fields using Eq. (\ref{eq23}) with $\mathbf{G}_{00} (\mathbf{x} -
\mathbf{x}')$. The $\boldsymbol{\gamma}''_{\alpha}$ are
\begin{equation}
\label{eqA12}
\begin{aligned}
  \boldsymbol{\gamma}''_{1} = \frac{1}{4\pi} \bigg[ & - \frac{( \lambda +
    3\mu)}{\mu ( \lambda + 2 \mu)} \ln \frac{r}{r_{0}} + \frac{(\lambda + \mu
    ) \, x^{2}}{ \mu ( \lambda + 2\mu )(x^{2}+y^{2})} \, ,
  \\
  & \frac{( \lambda + \mu ) \, xy}{ \mu ( \lambda + 2\mu ) (x^{2}+y^{2})}
  \, , \, 0  , \, 0 , \, 0  , \, 0 \,  \bigg]^{t} \, ,\\
  \boldsymbol{\gamma}''_{2} = \frac{1}{4\pi} \bigg[ & \, \frac{( \lambda + \mu
    ) \, xy}{ \mu ( \lambda + 2\mu ) (x^{2}+y^{2})} \, , \,
  - \frac{( \lambda + 3\mu )}{\mu ( \lambda + 2\mu )} \ln \frac{r}{r_{0}} \\
  & + \frac{( \lambda + \mu ) \, y^{2}}{ \mu ( \lambda + 2\mu )(x^{2}+y^{2})}
  \, , \, 0 , \, 0 , \, 0 , \, 0 \, \bigg]^{t} \, ,
  \\
  \boldsymbol{\gamma}''_{3} = \frac{1}{2\pi} \bigg[ & \, 0 , \, 0 , \, -
  \frac{1}{\mu} \ln \frac{r}{r_{0}} \, , \, 0 , \, 0 , \, 0 \, \bigg]^{t} \, ,
  \\
  \boldsymbol{\gamma}''_{4} = \frac{1}{2\pi} \bigg[ & \, 0 , \, 0 , \, 0 , \,
  - \frac{1}{\mu_{5}} \ln \frac{r}{r_{0}} \, , \, 0 , \, 0 \bigg]^{t} \, ,
  \\
  \boldsymbol{\gamma}''_{5} = \frac{1}{2\pi} \bigg[ & \, 0 , \, 0 , \, 0 , \,
  0 , \, - \frac{1}{\mu_{5}} \ln \frac{r}{r_{0}} \, , \, 0 \bigg]^{t} \, ,
  \\
  \boldsymbol{\gamma}''_{6} = \frac{1}{2\pi} \bigg[ & \, 0 , \, 0 , \, 0 , \,
  0 , \, 0 , \, - \frac{1}{\mu_{5}} \ln \frac{r}{r_{0}} \, \bigg]^{t} \,
  .  \\
\end{aligned}
\end{equation}
These fields don't have the unit of a length, but our algorithm multiplies them
by appropriate factors to become displacement fields.

\section*{Appendix C: Elastic constants}

Different notations concerning the five independent elastic constants
of icosahedral quasicrystals are in use.

To every elastic constant belongs a quadratic invariant of strain components.
Different sets of elastic constants can be compared with each other, when the
respective invariants are formulated in one and the same coordinate system.
This can be achieved by means of coordinate transformations. Relations between
different elastic constants arise from expressing one set of invariants by
another and comparing the coefficients.

All authors we adress also use the two Lam\'{e}-constants $\lambda$
and $\mu$ (\ref{eq15}), but a different phonon-phason-coupling and
other phasonic elastic constants.

In \cite{J11} and \cite{J26}, the phonon-phason-coupling is denoted
$R$, and the two phasonic elastic constants are $K_{1}$ and $K_{2}$.
The coordinate systems in phonon and phason spaces in these papers are
transformed into our fivefold $K_{5}^{u,w}$ when
substituting $x \rightarrow -x$, $y \rightarrow -y$, $z \rightarrow z$
and vice versa.  Transforming our elastic tensor $\mathbf{C}_{5}$ in
this manner allows a comparison with the parts $\mathbf{R}$ and
$\mathbf{K}$ of the elastic tensor given explicitly in \cite{J11}. The
relations between the elastic constants are
\begin{align}
  \mu_{3} & = \sqrt{6} \, R \, , & R & = \frac{1}{\sqrt{6}} \,
  \mu_{3} \, , \nonumber \\
  \label{eqA21}
  \mu_{4} & = K_{1} -2 K_{2} \, , &
  K_{1} & = \frac{1}{3} ( \mu_{4} + 2 \mu_{5}) \, ,\\
  \mu_{5} & = K_{1} + K_{2} \, , \hspace{6ex} & K_{2} & = -
  \frac{1}{3} ( \mu_{4} - \mu_{5}) \, . \nonumber
\end{align}

The phonon-phason-coupling of \cite{J27} is denoted $K_{3}$ and the phasonic
elastic constants are $K_{1}$ and $K_{2}$, but with another meaning as in
\cite{J11} and \cite{J26}. To compare the elastic energy expression of
\cite{J27} with our Eq. (\ref{eq14}), one must substitute
$\boldsymbol{\varepsilon}^{w}_{ij}$ by $\boldsymbol{\varepsilon}^{w}_{ji}$ for
all $i,j$. This is because the definition $\boldsymbol{\varepsilon}^{w}_{ij} =
\frac{\partial w_{j}}{\partial x_{i}}$ in \cite{J27} is different from our
definition (\ref{eq8}). In a second step, this new expression must be
transformed into our coordinate systems of Figs.  \ref{fig:1}, \ref{fig:2},
what is achieved by subjecting all components
$\boldsymbol{\varepsilon}^{u,w}_{ij}$ to the coordinate transformation $x
\rightarrow y$, $y \rightarrow -x$, $z \rightarrow z$. From the elastic energy
density in this new form, we see that the elastic constants are related by
\begin{align}
  \mu_{3} & = - \sqrt{6} \, K_{3} \, , &
  K_{3} & = - \frac{1}{\sqrt{6}} \, \mu_{3} \, , \nonumber \\
  \label{eqA23}
  \mu_{4} & = K_{1} + \frac{5}{3} K_{2} \, , & K_{1} & = \frac{1}{9}
  ( 4 \mu_{4} + 5 \mu_{5}) \, , \\
  \mu_{5} & = K_{1} - \frac{4}{3} K_{2} \, , \hspace{6ex} & K_{2} & =
  \frac{1}{3} ( \mu_{4} - \mu_{5}) \, . \nonumber
\end{align}

\end{document}